\documentclass[12pt]{article}
\usepackage{amssymb,amsfonts}
\usepackage{graphics,psboxit,amsmath}
\usepackage{subfigure}
\usepackage{graphicx}
\usepackage{verbatim}
\usepackage{mathrsfs}

\def\hybrid{\topmargin 0pt    \oddsidemargin 0pt 
        \headheight 0pt \headsep 0pt
        \textwidth 6.25in       
        \textheight 22.5cm       
        \marginparwidth .875in
        \parskip 5pt plus 1pt   \jot = 1.5ex}

\hybrid

\def\baselinestretch{1.2}

\catcode`\@=11

\def\marginnote#1{}
\newcount\hour
\newcount\minute
\newtoks\amorpm
\hour=\time\divide\hour by60 \minute=\time{\multiply\hour by60
\global\advance\minute by-\hour}
\edef\standardtime{{\ifnum\hour<12 \global\amorpm={am}%
        \else\global\amorpm={pm}\advance\hour by-12 \fi
        \ifnum\hour=0 \hour=12 \fi
        \number\hour:\ifnum\minute<10 0\fi\number\minute\the\amorpm}}
\edef\militarytime{\number\hour:\ifnum\minute<10 0\fi\number\minute}

\def\draftlabel#1{{\@bsphack\if@filesw {\let\thepage\relax
   \xdef\@gtempa{\write\@auxout{\string
      \newlabel{#1}{{\@currentlabel}{\thepage}}}}}\@gtempa
   \if@nobreak \ifvmode\nobreak\fi\fi\fi\@esphack}
        \gdef\@eqnlabel{#1}}
\def\@eqnlabel{}
\def\@vacuum{}
\def\draftmarginnote#1{\marginpar{\raggedright\scriptsize\tt#1}}

\def\draft{\oddsidemargin -.5truein
        \def\@oddfoot{\sl preliminary draft \hfil
        \rm\thepage\hfil\sl\today\quad\militarytime}
        \let\@evenfoot\@oddfoot \overfullrule 3pt
        \let\label=\draftlabel
        \let\marginnote=\draftmarginnote
   \def\@eqnnum{(\theequation)\rlap{\kern\marginparsep\tt\@eqnlabel}%
\global\let\@eqnlabel\@vacuum}  }

\def\draft2{
        \def\@oddfoot{\sl preliminary draft \hfil
        \rm\thepage\hfil\sl\today\quad\militarytime}
        \let\@evenfoot\@oddfoot \overfullrule 3pt
        \let\label=\draftlabel
        \let\marginnote=\draftmarginnote
   \def\@eqnnum{(\theequation)\rlap{\kern\marginparsep\tt\@eqnlabel}%
\global\let\@eqnlabel\@vacuum}  }

\def\preprint{\twocolumn\sloppy\flushbottom\parindent 2em
        \leftmargini 2em\leftmarginv .5em\leftmarginvi .5em
        \oddsidemargin -.5in    \evensidemargin -.5in
        \columnsep .4in \footheight 0pt
        \textwidth 10.in        \topmargin  -.4in
        \headheight 12pt \topskip .4in
        \textheight 6.9in \footskip 0pt
        \def\@oddhead{\thepage\hfil\addtocounter{page}{1}\thepage}
        \let\@evenhead\@oddhead \def\@oddfoot{} \def\@evenfoot{} }

\def\numberbysection{\@addtoreset{equation}{section}
        \def\theequation{\thesection.\arabic{equation}}}

\def\underline#1{\relax\ifmmode\@@underline#1\else
        $\@@underline{\hbox{#1}}$\relax\fi}

\def\titlepage{\@restonecolfalse\if@twocolumn\@restonecoltrue\onecolumn
     \else \newpage \fi \thispagestyle{empty}\c@page\z@
        \def\thefootnote{\fnsymbol{footnote}} }

\def\endtitlepage{\if@restonecol\twocolumn \else \newpage \fi
        \def\thefootnote{\arabic{footnote}}
        \setcounter{footnote}{0}}  

\catcode`@=12 \relax

\def\figcap{\section*{Figure Captions\markboth
        {FIGURECAPTIONS}{FIGURECAPTIONS}}\list
        {Figure \arabic{enumi}:\hfill}{\settowidth\labelwidth{Figure
999:}
        \leftmargin\labelwidth
        \advance\leftmargin\labelsep\usecounter{enumi}}}
 \relax
\def\tablecap{\section*{Table Captions\markboth
        {TABLECAPTIONS}{TABLECAPTIONS}}\list
        {Table \arabic{enumi}:\hfill}{\settowidth\labelwidth{Table
999:}
        \leftmargin\labelwidth
        \advance\leftmargin\labelsep\usecounter{enumi}}}
 \relax
\def\reflist{\section*{References\markboth
        {REFLIST}{REFLIST}}\list
        {[\arabic{enumi}]\hfill}{\settowidth\labelwidth{[999]}
        \leftmargin\labelwidth
        \advance\leftmargin\labelsep\usecounter{enumi}}}
 \relax
\makeatletter
\newcounter{pubctr}
\def\publist{\@ifnextchar[{\@publist}{\@@publist}}
\def\@publist[#1]{\list
        {[\arabic{pubctr}]\hfill}{\settowidth\labelwidth{[999]}
        \leftmargin\labelwidth
        \advance\leftmargin\labelsep
        \@nmbrlisttrue\def\@listctr{pubctr}
        \setcounter{pubctr}{#1}\addtocounter{pubctr}{-1}}}
\def\@@publist{\list
        {[\arabic{pubctr}]\hfill}{\settowidth\labelwidth{[999]}
        \leftmargin\labelwidth
        \advance\leftmargin\labelsep
        \@nmbrlisttrue\def\@listctr{pubctr}}}
 \relax
\makeatother

\def\be{\begin{equation}}
\def\ee{\end{equation}}
\def\ba{\begin{eqnarray}}
\def\ea{\end{eqnarray}}

\def\k{\kappa}
\def\r{\rho}
\def\a{\alpha}

\def\b{\beta}

\def\g{\gamma}

\def\d{\delta}
\def\D{\Delta}

\def\p{\pi}

\def\m{\mu}

\def\Om{\Omega}
\def\l{\lambda}
\def\L{\Lambda}
\def\s{\sigma}

\def\tp{\otimes}

\def\no{\noindent}

\def\qq{\qquad}

\def\IR{\relax{\rm I\kern-.18em R}}

\def \ov {\over}

\begin{document}


\renewcommand{\theequation}{\thesection.\arabic{equation}}
\csname @addtoreset\endcsname{equation}{section}

\newcommand{\eqn}[1]{(\ref{#1})}
\begin{titlepage}
\begin{center}

\strut\hfill   LAPTH-1367/09
\vskip 1.0cm


{\Large \bf Introduction to Quantum Integrability}

\vskip 0.5in

{\bf Anastasia Doikou$^{a}$, Stefano Evangelisti$^{b}$, Giovanni Feverati$^{c}$ {\footnotesize and}\\
Nikos Karaiskos$^{a}$}
\footnote{\tt This article is based on a series of lectures presented
at the University of Bologna in November 2007 by A.D. and G.F., and University of Patras in May 2009 by A.D.}

\vskip 0.1in

{\footnotesize $^{a}$ Department of Engineering Sciences, University of Patras,\\
26110 Patras, Greece}\phantom{x}

{\footnotesize $^{b}$ University of Bologna, Physics Department, INFN-Sezione di Bologna\\
Via Irnerio 46, 40126 Bologna, Italy}

{\footnotesize $^{c}$ Laboratoire de Physique Theorique, LAPTH
CNRS, UMR 5108,\\
F-74941, Annecy-le-Vieux, France}

\vskip .2in


{\footnotesize {\tt adoikou$@$upatras.gr, stefano.evangelisti@gmail.com,
feverati@lapp.in2p3.fr, nkaraiskos@upatras.gr}}\\

\end{center}

\vskip 0.5in

\centerline{\bf Abstract}
In this article we review the basic concepts regarding quantum integrability. Special emphasis is given
on the algebraic content of integrable models. The associated algebras are essentially described by the
Yang-Baxter and boundary Yang-Baxter equations depending on the choice of boundary conditions. The relation between
the aforementioned equations and the braid group is briefly discussed.
A short review on quantum groups as well as the quantum inverse scattering method
(algebraic Bethe ansatz) is also presented.
\no

\vfill

\end{titlepage}
\vfill \eject


\tableofcontents

\def\baselinestretch{1.2}
\baselineskip 20 pt
\no

\section{Introduction}

The main purpose of this article is to offer a review on the basic ideas
of quantum integrability as well as familiarize the reader, who has not necessarily a background on the subject,
with the fundamental concepts.

Quantum integrability in 1+1 dimensions has been a very rich research subject, especially after the seminal works
of the St. Petersburg group (see e.g. \cite{FTS}--\cite{tak}) on the quantum inverse scattering method (QISM).
We refer the interested reader to a number of lecture notes and review articles on algebraic Bethe ansatz, special topics
on integrable models,
or articles with emphasis on statistical and thermodynamic properties or applications to condensed matter physics
(see e.g \cite{faddeev1}--\cite{klumper}).
In these notes we are basically focusing on the algebraic content of quantum integrable systems
giving particular emphasis
on the quantum algebras and their connections to braid groups and Hecke algebras. We also review the quantum inverse
scattering method and
briefly discuss lattice integrable
models with open boundary conditions.

The outline of the article is as follows: in the next section we introduce the basic notation on tensor products of
matrices and vectors and we briefly review the $\mathfrak{su}_2$ algebra as well as its representations.
We then introduce
the Heisenberg model \cite{bethe}, describing first neighbors spin-spin interaction. In section 3 we present
in more detail the XXX
(isotropic)
and XXZ (anisotropic Heisenberg) models.
In particular, we give a first flavor on the corresponding spectra and eigenstates for small a number of sites.
We also discuss the
zero temperature phase diagram.
The next section is basically devoted to the Yang-Baxter \cite{baxter} equation and its solution, the so called $R$ matrix.
This is the fundamental equation within
the QISM context. We introduce the equation and also provide systematic means for solving it via its structural
similarity with the
braid group. The braid group and certain quotients, such as the Hecke and Temperley-Lieb algebras \cite{hecke, hecke1, tl},
are also discussed.

In section 5 we introduce the quantum Lax operator, and the fundamental algebraic equation governing the
underlying quantum algebras (Yangians and $q$ deformed Lie algebras) \cite{drinf, jimbo}.
We then construct tensorial representations of the underlying algebras, and eventually build the closed (periodic)
transfer matrix
of a spin chain-like system. We show the integrability of the system, and also extract the corresponding local Hamiltonian.
In the next section we discuss in more detail the non-trivial co products arising in quantum algebras and we show
how one can exploit them in order to investigate
the symmetry of the associated $R$ matrix. In section 7 we present representations of the
$U_q(\mathfrak{sl}_2)$ algebra and
discuss in detail the algebraic Bethe ansatz technique for diagonalizing the
generalized XXZ spin chain. In the last section we
discuss integrable lattice models with generic integrable boundary conditions \cite{sklyanin}. The corresponding
fundamental algebraic relation
i.e. the reflection equation \cite{cherednik} is introduced and  solutions (reflection matrices)
are obtained with the help
of the $B$-type braid group and its quotients \cite{cher, male, mawo, doma}. Tensorial representations
of the reflection algebra are
constructed and the open transfer matrix is
introduced. Finally, the $U_q(\mathfrak{sl}_2)$ invariant open XXZ spin chain \cite{kusk1} is discussed
and the corresponding quadratic Casimir is extracted from
the open transfer matrix.



\section{Preliminaries}

\subsection{Notation}

Before we proceed with the presentation of the fundamental notions of quantum integrability
it is necessary to introduce some basic notation.

Consider the tensor vector space $V \otimes V$ then define
\ba
A_1 & = & A \tp \mathbb{I}\cr
B_2 & = & \mathbb{I} \tp B. \label{notation}
\ea
We then attach subscripts on the various elements to define the
respective vector space on which they act non-trivially. For example, suppose
that $A,\ B \in \mbox{End}V$. In the described notation the tensor product between them
can be written as
\be
A \tp B = A_1\ B_2.
\ee
In general, consider the tensor sequence of $N$ vector spaces $V \otimes V \otimes \ldots \otimes V$ then define:
\be\label{tensorindex}
A_n = {\mathbb I} \otimes  \ldots \otimes {\mathbb I} \otimes \underbrace{A}_n \otimes
{\mathbb I} \otimes  \ldots \otimes {\mathbb I}, ~~~~~n \in \{1,\ 2, \ldots, N \}.
\ee
We shall extensively use such notation subsequently
when constructing one dimensional
integrable quantum spin chains, which is one of the
primary objectives of this review.

Some basic properties of the tensor product are listed below\footnote{Note that for
the first of the properties listed we are focusing on
non super symmetric algebras. In the super symmetric case it  is modified in accordance to
the fermionic and bosonic degrees of freedom.}:
\ba
&&(A \otimes B)\ (C \otimes D)= A C \otimes B D\cr
&& (A\otimes B)^{-1} = A^{-1} \otimes B^{-1} \cr
&& (A \otimes B)^{T} = A^T \otimes B^T.
\ea
There is a simple rule that gives the tensor product of two matrices.
Consider for simplicity the $2 \times 2$ matrices
$A= \left( \begin{array}{cc}
	               a_{11}  & a_{12} \\
				   a_{21}  & a_{22}
\end{array} \right)$ and
$B= \left( \begin{array}{cc}
	               b_{11}  & b_{12} \\
				   b_{21}  & b_{22}
\end{array} \right)$, the tensor product $A \otimes B$ is a $4\times 4$ matrix defined
as
\ba
A \otimes B &=&
\left( \begin{array}{cc}
	               a_{11}B  & a_{12}B \\
				   a_{21}B  & a_{22}B
\end{array} \right) \cr &=& \left( \begin{array}{cc|cc}
a_{11} b_{11} & a_{11} b_{12} & a_{12} b_{11} & a_{12} b_{12}  \\
a_{11} b_{21} & a_{11} b_{22} & a_{12} b_{21} & a_{12} b_{22}  \\
\hline
a_{21} b_{11} & a_{21} b_{12} & a_{22} b_{11} & a_{22} b_{12}  \\
a_{21} b_{21} & a_{21} b_{22} & a_{22} b_{21} & a_{22} b_{22}
\end{array} \right) \,.
\label{tensormatrix}
\ea
In general for two $n \times n$ matrices $A,\ B$ the corresponding tensor product
is a $n^2 \times n^2$
matrix and the rule
generalizes in a straightforward manner: $(A \otimes B)_{ij, kl}= a_{ij}\ b_{kl}$.

The tensor product of two vectors $a,\ b \in {\mathbb C}^2$ is derived as
\ba
a \otimes b = \left( \begin{array}{c}
      a_{1} \\
      a_{2}
\end{array} \right) \otimes
\left( \begin{array}{c}
      b_{1} \\
      b_{2}
\end{array} \right) =
\left( \begin{array}{c}
      a_{1} b_{1} \\
      a_{1} b_{2} \\
	  \hline
	  a_{2} b_{1} \\
	  a_{2} b_{2}
\end{array} \right) \,,
\label{tensorvectors}
\ea
and in general for $n$ column vectors $a,\ b \in {\mathbb C}^n$ we obtain an $n^2$ column vector:
$(a \otimes b)_{i,j} = a_i\ b_j$.


\subsection{The $\mathfrak{su}_2$ algebra: a brief review}

The $\mathfrak{su}_2$ algebra is defined by the generators $J^{\pm},\ J^z$ and
the exchange relations
\ba
[J^+,\ J^-] & = & 2 J^z\cr
[J^z,\ J^{\pm}] & = & \pm J^{\pm}.
\label{rels}
\ea
\no
The spin ${1 \over 2}$ representation of $\mathfrak{su}_2$  maps the three
generators of the algebra to the three Pauli matrices. Indeed consider the spin ${1\over 2}$
representation
$\pi:\ \mathfrak{su}_2\ \hookrightarrow \mbox{End}({\mathbb C}^2)$ such that:
\be
\pi(J^z) = {1\over 2}\sigma^z, ~~~~~\pi(J^{\pm}) = \sigma^{\pm}
\ee
and $\sigma^{\pm},\ \sigma^z$ are the familiar $2 \times 2$ Pauli matrices
\be
\sigma^z=\begin{pmatrix}1 & 0 \\ 0 & -1\end{pmatrix}, \qquad \sigma
^+=\begin{pmatrix} 0 & 1 \\ 0 & 0 \end{pmatrix},
\qquad \sigma^-=\begin{pmatrix}0 & 0 \\ 1 & 0 \end{pmatrix}.
\label{pauli}
\ee
There is a homomorphism $h$ defined as $h: \mathfrak{su}_2 \hookrightarrow \mathfrak{sl}_2$ such that
\be
h(J^z) = J^z\ , \qq
h(J^+) = J^+\ , \qq h(J^-) =-J^-\ ,
\ee
so the two algebras $\mathfrak{su}_2$, $\mathfrak{sl}_2$ will be equivalently used henceforth.

The typical spin $s$ representation of $\mathfrak{su}_2$ is  an $n =2s +1$ dimensional
representation and may be expressed in terms of $n \times n$ matrices; infinite
dimensional representations exist
in terms of differential operators. Define first the $n \times n$ matrices
$e_{ij}$ such that
\be
(e_{ij})_{kl} = \delta_{ik} \delta_{jl}, ~~~~~
\mbox{and} ~~~~~ e_{ij}\ e_{kl} = \delta_{jk}\ e_{il}.
\ee
Consider the $n$ dimensional matrix representation: $\rho:\ \mathfrak{su}_2\ \hookrightarrow\ \mbox{End}
({\mathbb C}^n)$
such that:
\be
\rho(J^z) = \sum_{k=1}^n \alpha_k\ e_{kk}, ~~~~\rho(J^+) = \sum_{k=1}^{n-1} C_k\ e_{k k+1},
~~~~\rho(J^-) = \sum_{k=1}^{n-1} C_k\ e_{k+1 k} \label{rep11}
\ee
where we define
\be
 \alpha_k = {n+1 \over 2} -k, ~~~~C_k = \sqrt{k(n-k)}. \label{r1}
\ee

The generic spin $s$ representation of $\mathfrak{sl}_2$  in terms of differential operators may be expressed as:
\be
J^z \hookrightarrow y {d \over d y}, ~~~~~J^+ \hookrightarrow y^{-1}(y {d \over d y} + s), ~~~~~J^- \hookrightarrow
y(y {d \over d y} - s).
\label{r2}
\ee
In this case $s$ may be any number --not necessarily an integer-- and the space of
functions is infinite dimensional.
It is a straightforward exercise to show
that
(\ref{rep11}), (\ref{r2}) satisfy the $\mathfrak{sl}_2$ exchange relations.


\subsection{The Heisenberg model}

We come now to the description of physical models associated to the $\mathfrak{su}_2$ algebra described above.
The Heisenberg model was introduced as a natural physical description
of magnetism in solid state physics \cite{bethe}.
Its one-dimensional version, the only one of interest here, has also the merit of
having inaugurated the studies of quantum integrable systems and of
the methods known as Bethe Ansatz \cite{bethe, FT, FT2}.

The Heisenberg idea is to consider, on each lattice site, a quantum magnetic
needle of spin $\frac{1}{2}$, fully free to rotate.
Formally, this is represented by a two-dimensional local space of states
$\mathbb{C}^2$ that can accommodate a spin up and a spin down components
\begin{equation}
\label{YB3}
\mid \uparrow \rangle =\binom{1}{0} \qquad \mid \downarrow \rangle =\binom{0}{1}
\end{equation}
The full space of states is then consisting of sums of tensor products of such spins up and down on
all sites of the lattice, here taken one dimensional and consisting of $N$ sites:
\be
\label{YB2}
W=\mathbb{C}^2\otimes\mathbb{C}^2\otimes\dots\otimes\mathbb{C}^2
\ee
The local spin operators can be introduced by using Pauli matrices on each site
\be\label{pauli2}
S_i^{z}=\frac12 \sigma_i^{z}, ~~~~~S_i^{x,y} = \frac12 \sigma_i^{x,y}
\ee
where the Pauli matrices act on the $i$-th site
of the chain, according to \ref{tensorindex}.
Recall that
\be
\sigma^+ ={1\over 2} (\sigma^x +i \sigma^y), ~~~~~\sigma^- = {1\over 2}(\sigma^x -i \sigma^y).
\label{relations}
\ee
The Pauli matrices are given by (\ref{pauli}), and clearly satisfy the $\mathfrak{su}_2$
commutation relations. Following (\ref{tensorindex}), Pauli matrices acting on different sites commute
with each other:
\be
[\sigma_i^{\xi},\ \sigma_j^{\xi'}]=0\,,\quad i\neq j, ~~~\xi \in \{x,\ y,\ z\}. \label{different}
\ee

The magnetic needle is assumed sensitive to the nearest neighbor needles
with the simplest possible coupling of magnetic dipoles
\be
{\mathrm J}_x S_i^x S_{i+1}^x+{\mathrm J}_y S_i^y S_{i+1}^y +{\mathrm J}_z S_i^z S_{i+1}^z,
\ee
then the corresponding Hamiltonian is obtained by summing on all the lattice sites
\be\label{fullh}
H=\sum_i\left({\mathrm J}_x S_i^x S_{i+1}^x+{\mathrm J}_y S_i^y S_{i+1}^y +{\mathrm J}_z S_i^z S_{i+1}^z\right).
\ee
Here the couplings ${\mathrm J}_{\xi}$ are taken constant through the lattice.
This model is known as XYZ model.

Often, for computational reasons, it is convenient to impose some special
conditions at the two edges of the lattice. These can also represent the
interaction of the lattice with some environment.
A very common choice, that will be adopted here, is to consider
the first and last sites to be adjacent
\be\label{pbc}
N+1\equiv 1 \quad \Rightarrow \quad i + N \equiv i
\ee
this is know as periodic boundary conditions.

\begin{figure}
\begin{center}
\includegraphics[height=25mm]{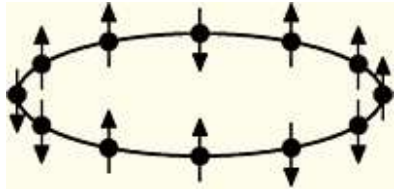}
\caption{Spin chain with periodic boundary conditions.}
\label{sc}
\end{center}
\end{figure}

These conditions `close' the
chain (see Figure 1) and are a bit unphysical\footnote{These is no physical justification
for the assumption that, in a piece of material, the atoms
at opposite extremes should be considered as interacting. In spite of that,
thermodynamic properties are independent of boundary conditions if
interactions are short ranged therefore periodic boundary conditions can be used
to obtain realistic results.}, but
often extremely convenient; more physical are the open boundaries conditions,
where no interaction between the first and the last sites is assumed (see last lecture).

The Hamiltonian (\ref{fullh}) with constant coefficients can be exactly treated
with functional methods. Indeed, this model is equivalent to the
8-vertex model solved by Baxter \cite{baxter}. Here a different and simpler
approach will be used --algebraic Bethe ansatz-- that works for the so called XXX and XXZ models,
obtained by identifying the coefficients ${\mathrm J}_x = {\mathrm J}_y = {\mathrm J}_z$
and ${\mathrm J}_x={\mathrm J}_y$ respectively. Note that the XYZ model can be also solved via
the algebraic Bethe ansatz
methodology, but after implementing certain modifications that involve specific local gauge transformations
(see \cite{FT2}). We shall not
however discuss these subtle technical points here.

\section{The XXX and XXZ quantum spin chains}

\subsection{The XXX model}

Our main purpose now is to diagonalize local Hamiltonians describing
interactions between first neighbors
as described by (\ref{fullh}).
We shall first introduce the 2-site XXX or isotropic Heisenberg model.
For the two-site problem, the basic observables are the spin operators
at each site, $\vec \sigma_{1} \equiv \vec \sigma \otimes \mathbb{I}$
and $\vec \sigma_{2} \equiv \mathbb{I} \otimes \vec \sigma$.
That is (see also \cite{nepomechie}),
\ba
\sigma_{1}^{x} &=& \left( \begin{array}{cc}
	               0  & 1 \\
				   1  & 0
\end{array} \right) \otimes \left( \begin{array}{cc}
	               1  & 0 \\
				   0  & 1
\end{array} \right) = \left( \begin{array}{cc|cc}
 &  &  1  &0 \\
 &  & 0 &  1  \\
\hline
 1  & 0 &  & \\
 0&  1  &  &
\end{array} \right) \,, \cr
\sigma_{1}^{y} &=& \left( \begin{array}{cc}
	               0  & -i \\
				   i  & 0
\end{array} \right) \otimes \left( \begin{array}{cc}
	               1  & 0 \\
				   0  & 1
\end{array} \right) = \left( \begin{array}{cc|cc}
 &  &  -i &0  \\
 &  &  0& -i \\
\hline
 i  &0  &  & \\
 0&  i  &  &
\end{array} \right) \,, \cr
\sigma_{1}^{z} &=& \left( \begin{array}{cc}
	               1  & 0 \\
				   0  & -1
\end{array} \right) \otimes \left( \begin{array}{cc}
	               1  & 0 \\
				   0  & 1
\end{array} \right) = \left( \begin{array}{cc|cc}
 1 &0  &  &  \\
 0& 1  &  &  \\
\hline
 & & -1  & 0\\
 & &  0& -1
\end{array} \right) \,,
\ea
and
\ba
\sigma_{2}^{x} &=& \left( \begin{array}{cc}
	               1  & 0 \\
				   0  & 1
\end{array} \right) \otimes
\left( \begin{array}{cc}
	               0  & 1 \\
				   1  & 0
\end{array} \right)   = \left( \begin{array}{cc|cc}
 0&  1  &  & \\
 1  &0  &  & \\
\hline
 &  &  0& 1  \\
 &  &  1 &0
\end{array} \right) \,, \cr
\sigma_{2}^{y} &=& \left( \begin{array}{cc}
	               1  & 0 \\
				   0  & 1
\end{array} \right)  \otimes
\left( \begin{array}{cc}
	               0  & -i \\
				   i  & 0
\end{array} \right) = \left( \begin{array}{cc|cc}
 0& -i  &  & \\
 i  &0  &  & \\
\hline
 & & 0& -i   \\
 & & i  &0
\end{array} \right) \,, \cr
\sigma_{2}^{z} &=& \left( \begin{array}{cc}
	               1  & 0 \\
				   0  & 1
\end{array} \right) \otimes \left( \begin{array}{cc}
	               1  & 0 \\
				   0  & -1
\end{array} \right) = \left( \begin{array}{cc|cc}
 1  &0  &  & \\
0 & -1  &  & \\
\hline
 &  &  1 &0  \\
 &  &  0& -1
\end{array} \right) \,.
\ea
Recall also (\ref{different}).
These operators act on the tensor product space ${\mathbb C}^2 \otimes {\mathbb C}^2$, with
elements $a \otimes b$ given explicitly in (\ref{tensorvectors})

The 2-site Hamiltonian for the XXX model is then given by
\be\label{hamilxxx}
H_{12}=-{1\ov 2}(\s_1^x\s_2^x+\s_1^y\s_2^y+\s_1^z\s_2^z),
\ee
\no
which describes the interaction of two spin 1/2 magnets.

It is easy to verify that the 2-site Hamiltonian
may be alternatively written as:
\be\label{duesiti}
H_{12} =- \mathcal{P} + c\ \mathbb{I},
\ee
\no where
\be
\mathcal{P} =
\begin{pmatrix}
	1 & 0 & 0 & 0 \\
	0 & 0 & 1 & 0 \\
	0 & 1 & 0 & 0 \\
	0 & 0 & 0 & 1
\end{pmatrix} = {1\over 2} (\sigma^x \otimes \sigma^x +\sigma^y \otimes \sigma^y +
\sigma^z \otimes \sigma^z +{\mathbb I} \otimes {\mathbb I}), \label{perm}
\ee
\no is the \emph{permutation operator}. The constant here is $c=\frac12$.

\paragraph{Exercise 1:} Show that $\mathcal{P}$
satisfies the following properties:
\ba
\mathcal{P}\ (a\tp b) & = &  b\tp a, \qquad \forall\ a,b \in {\mathbb C}^2,\\
\mathcal{P}_{12}\ A_1\ B_2\ \mathcal{P}_{12} & = & A_2\ B_1,\\
\mathcal{P}^2_{12} & = & \mathbb{I}.
\ea
where $A,\ B$ are $2\times 2$ matrices.

One can then generalize the permutation operator
$\mathcal{P}$ acting on $\mathbb{C}^n \otimes {\mathbb C}^n$, and defined as
\be
\mathcal{P}=\sum_{i,j=1}^n e_{ij}\tp e_{ji}.
\ee
\paragraph{Exercise 2:} Show explicitly that $\mathcal{P}^2=\mathbb{I}$
holds for the generic case. Do the same for the properties of the
permutation operator
\ba
\mathcal{P}_{12}\ A_1\ \mathcal{P}_{12}  =  A_2\ \Leftrightarrow\
\mathcal{P}_{12}\ A_1  =  A_2\ \mathcal{P}_{12}.
\ea
where $A$ is any $n \times n$ matrix that can be expressed as: $A= \sum_{i,j =1}^n a_{ij}e_{ij}$.

\paragraph{Exercise 3:} Show by inspection that the following states
\ba
|\psi_1\rangle & = & |\uparrow\rangle\tp |\uparrow\rangle, \qquad |\psi_2\rangle = |\downarrow\rangle\tp |\downarrow\rangle\cr
|\psi_3\rangle & = & |\uparrow\rangle\tp |\downarrow\rangle + |\downarrow\rangle\tp |\uparrow\rangle\cr
|\psi_4\rangle & = & |\uparrow\rangle\tp |\downarrow\rangle - |\downarrow\rangle\tp |\uparrow\rangle,
\ea
\no where $|\uparrow\rangle,\ |\downarrow\rangle$ are defined in (\ref{YB3}), are eigenfunctions of the Hamiltonian. Find also
the corresponding eigenvalues and discuss the possible degeneracies.

\paragraph{Exercise 4:} Define the \emph{co-product} of $\mathfrak{sl}_2$ as a mapping
$\Delta:\mathfrak{sl}_2 \hookrightarrow \mathfrak{sl}_2\otimes \mathfrak{sl}_2$
\be
\Delta (X)= X_1 + X_2 = X \otimes \mathbb{I} + \mathbb{I} \otimes X, ~~~~~X \ \in \mathfrak{sl}_2.
\ee
\no
Show that $\Delta(J^{\pm}),\ \Delta(J^z)$, satisfy the relations of the $\mathfrak{sl}_2$
algebra. In other words, show that the co-product is a tensor realization of
the algebra.

\paragraph{Exercise 5:} Show that the 2-site Hamiltonian enjoys the $\mathfrak{su}_2$ symmetry,
that is:
\be
[X_1 + X_2,\ H_{12}]=0,
\ee
\no where $X\ \in \{\sigma^{\pm},\ \sigma^z\}$.
\newline
\\
{\bf The N-site Hamiltonian}
\\
The Hamiltonian can be generalized to the case of $N$ `particles', described
by the $N$-site Hamiltonian:
\be
H=\sum_{i=1}^{N}H_{i\,i+1}=
-{1\over 2}\sum_{i=1}^{N}(\s_i^x\s_{i+1}^x+\s_i^y\s_{i+1}^y+\s_i^z\s_{i+1}^z), \label{nham}
\ee
\no which clearly acts on $W=({\mathbb C}^2)^{\otimes N}$,
and $\dim W=2^N$. The model is
integrable and may be exactly solved \cite{bethe}.
In general the problem of diagonalizing the
Hamiltonian is quite tedious, even
with the use of numerical methods. However, one can find analytical solutions to this problem
by using quite sophisticated and powerful techniques under the name of Bethe ansatz
(see e.g. \cite{FTS, FT, faddeev1, faddeev2, sklyaninl}),
which will be one of the main subjects of the following sections. Here we shall focus on one of the many
variations of the
Bethe ansatz method that is the algebraic Bethe ansatz \cite{FTS, FT}.

The Hamiltonian (\ref{nham}) is manifestly translation invariant, made evident
by shifting $i$ into $i+1$ in the sum.
The translation operator corresponding to the translational symmetry is
\be
\label{trans}
\Pi = {\cal P}_{12}\ {\cal P}_{23} \ldots {\cal P}_{L-1,L}=e^{-i\mathbb{P}}
\ee
where ${\cal P}$ is the permutation operator; $\Pi$ commutes with the Hamiltonian (see exercise in Lecture 5).
The operator in the exponent is
the momentum operator, and is defined up to $2\pi$.
The action of $\Pi$ consists of shifting each site of the lattice by one step
to the right.

Before concluding, we should note that there also exists a generalization of the co-product
for the case of $N$ particles,
\be
\Delta^{(N)}(X)=\sum_{i=1}^N X_i=\sum_i\mathbb{I}\tp\mathbb{I}\tp\cdots\tp \underbrace{X}_i \tp\mathbb{I}
\tp\cdots\tp\mathbb{I}.
\ee

\paragraph{Exercise 6:} Show that the $N$-co-product commutes with the XXX Hamiltonian
\be
[H,\ \Delta^{(N)}(X)]=0, \qquad X\ \in \{\sigma^{\pm},\ \sigma^z\}
\ee
i.e. the XXX Hamiltonian is $\mathfrak{su}_2$ symmetric.

\subsection{The XXZ model}

We come now to the anisotropic model, that is the XXZ spin chain; we simply introduce
an anisotropy along $z$  by
\begin{equation}
\label{YB1}
H=-\frac{1}{2}\sum_{i=1}^{N}\left (\sigma_{i}^{x}\sigma_{i+1}^{x}+\sigma_{i}^{y}
\sigma_{i+1}^{y}+\Delta \sigma_{i}^{z}\sigma_{i+1}^{z}\right).
\end{equation}
This Hamiltonian acts on the Hilbert space defined in (\ref{YB2}).
The exact behavior of the system
depends on the coupling constant $\Delta$, that describes the anisotropy
of the model.

We shall consider periodic boundary conditions on the chain as in (\ref{pbc});
for local\footnote{A local operator acts on a single site only.} operators
this is expressed by
$\vec{\sigma}_{i+N} =\vec{\sigma}_{i}$ as is shown in
figure (\ref{sc}).
The XXZ Hamiltonian  is also translation invariant and the translation operator is given by (\ref{trans}).
The case $\Delta \neq 1$, or XXZ model was solved by Hans Bethe in 1931 \cite{bethe}, by using his famous
{\em ansatz}.
The case $\Delta = 1$ corresponds to the properly named Heisenberg model
or XXX model, discussed previously, which
as already mentioned (see exercise in previous section) has the global symmetry $\mathfrak{su}(2)$.
If $\Delta=0$ the $z$ component disappears and the model is called XY.
If $\Delta\rightarrow \pm \infty$, one actually obtains the Ising model.
One simple way to see this is to introduce a new Hamiltonian
by dividing by the coupling; the first group of terms disappears
\begin{equation}
\label{diviso}
H_{\text{Ising}}=\lim_{\Delta \pm \infty}\frac{H}{\Delta}
=-\frac{1}{2}\sum_{i=1}^{N} \sigma_{i}^{z}\sigma_{i+1}^{z},
\end{equation}
and the model is actually indistinguishable from the classical spins model
\be\label{ising}
-\frac{1}{2}\sum_{i=1}^{N} \tau_{i}\tau_{i+1}
\ee
with $\tau_i=\pm 1$, known as the Ising model.
For this reason, the XXZ model is indicated also as the
Heisenberg-Ising model.

By the $N$ co-product we can define the total spin operator ${\mathrm S}^{\xi}$
\begin{equation}
\label{YB10}
\Delta(S^{\xi})={\mathrm S}^{\xi}=\sum_{i=1}^{N} S_i^{\xi}.
\end{equation}
The third component is a symmetry of the model, no matter what the value of $\Delta$ is,
\be
\left [H,\ {\mathrm S}^{z}\right ] = 0.
\ee
To verify this last statement, the relations of $\mathfrak{su}_2$ (satisfied by Pauli matrices)
are essentially needed, plus the fact that
different spaces in the tensor product sequence commute with each other.

If $\Delta =1$, it is immediate to check, (see exercise in the previous section),
that the $x$ and $y$ components also commute therefore
\be
\left [H,\ {\mathrm S}^{\xi} \right ] = 0\,,\qquad \xi \in \{x,\,y,\,z\}
\ee
and the Hamiltonian is fully $\mathfrak{su}(2)$ symmetric.

Because of the indicated symmetry, in the XXZ model the third component ${\mathrm S}_z$
is particularly important; it will also be used as an order parameter,
because it specifies the different phases of the model itself.

The operator ${\mathrm S}_z$ is the sum of mutually commuting diagonal operators;
each component contributes to the eigenvalues with $\pm\frac12$ therefore
the spectrum is
\be\label{spectrumz}
-\frac{N}2,\ -\frac{N}2+1,\ -\frac{N}2+2,\ldots,\frac{N}2
\ee

There is a reflection symmetry
\be
\sigma^z_i\rightarrow -\sigma_i^z
\ee
that makes the spectrum of $H$ symmetric with respect to the values of ${\mathrm S}_z$.
\\
\\
{\bf Example: two-site Hamiltonian.}
\\
Consider now the two-site Hamitloanian
\be
H = - (\sigma^x \otimes \sigma^x + \sigma^y \otimes \sigma^y + \Delta \sigma^z \otimes \sigma^z)
\ee
The base vectors are ordered as
$\{\mid \uparrow\uparrow \rangle,\mid \uparrow\downarrow\rangle,
\mid \downarrow\uparrow \rangle,\mid \downarrow\downarrow\rangle\}$,
and for a two-sites lattice the Hamiltonian reads (the sum in (\ref{YB1}) goes over $j=1$ and $j=2$)
\begin{equation}
\label{YB9}
H=\left (
\begin{array}{cccc}
-\Delta & 0 & 0 & 0 \\ 0 & \Delta & -2 & 0 \\ 0 & -2 & \Delta & 0 \\ 0 & 0 & 0 & -\Delta
\end{array}\right).
\end{equation}
This matrix is symmetric and has real entries therefore is diagonalizable and
has real eigenvalues.
It has the following set of eigenvectors and eigenstates:
\be\label{autovalori}
\left \{\mid \uparrow\uparrow \rangle, -\Delta \right \}\,,
\left \{\mid \downarrow\downarrow \rangle, -\Delta \right \}\,,
\left \{\mid \uparrow\downarrow\rangle + \mid \downarrow\uparrow\rangle, \Delta -2 \right \}\,,
\left \{\mid \uparrow\downarrow\rangle -\mid \downarrow\uparrow\rangle, \Delta +2 \right \}
\ee
We examine the ground state to understand the possible phases of the system.

When $\Delta > 1$, the ground state has energy $E=-\Delta$ and it can be
one of the states $\mid \uparrow\uparrow \rangle,\,\mid \downarrow\downarrow\rangle$
therefore the system is ferromagnetic and the total magnetization is $M={\mathrm S}^z=2$.
This extends immediately to an arbitrary number of sites.

If $\Delta < 1$, the ground state is
$\mid \uparrow\downarrow\rangle + \mid \downarrow\uparrow\rangle$ with energy
$E=\Delta - 2$ therefore the system has vanishing total magnetization
$M={\mathrm S}^z=0$.
Notice that the Hamiltonian model is at zero temperature; the temperature has
not been introduced neither any statistical ensemble. With that in mind,
the ground state for $\Delta < 1$ is not a Gibbs mixture of states, but is
given as indicated: it has a global order (at least in the two-site case!)
and is actually called anti-ferromagnetic state\footnote{The regions $\Delta<-1$
and $-1<\Delta<1$ can be further discriminated if an external magnetic
field is added to the Hamiltonian}.

The doubly degenerate level $E=-\Delta$ becomes triply degenerate
if $\Delta = 1$ or
$\Delta = -1$. The case $\Delta=1$
is especially interesting, because the three coinciding eigenvalues $E=-\Delta=-1$
correspond also to the lowest energy. This marks a so called quantum phase transition,
namely a transition not induced by thermal fluctuations but by the variation of a parameter in the
Hamiltonian. It occurs at zero temperature only.

Some general comments are now in order.
The XXX Hamiltonian is free of couplings apart from the overall sign.
Given our sign choice, it is apparent that adjacent parallel spins
lower the energy. This explains the custom to call ferromagnetic
the Hamiltonian in (3.14) and anti-ferromagnetic the opposite one.
The presence of $\Delta$ in the XXZ model spoils this distinction
because the ferromagnetic/antiferromagnetic behavior depends on the
coupling and, as standard in physics, the name is attached to the
phase of the system. Here no temperature is introduced (namely we
are at zero temperature) so the phase is dictated by the ground state:
$\Delta >1$, the ground state is ferromagnetic;
$\Delta <1$, the ground state is anti-ferromagnetic;
$\Delta=1$, critical case, the ground state is multi-degenerate.

In exercises 8, 9, 10 below the reader can construct slightly larger cases, and
have a more accurate indication of the phases for the different values of
$\Delta$.

\paragraph{Exercise 7:} Prove that the Hamiltonian (\ref{YB1})
is real and symmetric therefore diagonalizable with real eigenvalues.

\paragraph{Exercise 8:} Construct the three-site Hamiltonian and check that the
ground-state for $\Delta>1$ is ferromagnetic while for $\Delta<1$ is anti-ferromagnetic
with frustration\footnote{Frustration indicates the phenomenon where adjacent magnets
tend to be antiparallel but geometry or topology forces a pair of them to be
parallel. In the present case, on a periodic odd sites lattice, an alternating sequence of
up and down spins is always frustrated. See later.}.

\paragraph{Exercise 9:} Construct the four-site Hamiltonian and check that the
groundstate for $\Delta>1$ is ferromagnetic. Find the ground state for $\Delta<1$, call it $\phi$.
This state is anti-ferromagnetic. Check that it is proportional to
\be\label{antif4}
\phi \propto
|\uparrow\uparrow\downarrow\downarrow\rangle
+|\uparrow\downarrow\downarrow\uparrow\rangle+|\downarrow\downarrow\uparrow\uparrow\rangle
+|\downarrow\uparrow\uparrow\downarrow\rangle+\frac{-\Delta\pm\sqrt{\Delta^2+8}}{2}
\big(|\uparrow\downarrow\uparrow\downarrow\rangle+
|\downarrow\uparrow\downarrow\uparrow\rangle\big)
\ee

\paragraph{Exercise 10:} In the special $\Delta=1$ case, the anti-ferromagnetic ground state
for a four-site lattice can be obtained just using symmetry considerations and the repeated
action of ${\mathrm S}^{-}$ on $|\uparrow\uparrow\uparrow\uparrow\rangle$. Find it and compare with
the longer method of Exercice 9.

\subsubsection{The Ising limits}

The argument given before (\ref{diviso}) is extremely useful to have an indication of
the phases.
For $\Delta\gg 1$, the Ising-like $z$ component of the hamiltonian
\be
\frac{H}{\Delta} =
-\frac{1}{2}\sum_{i=1}^{L} \sigma_{i}^{z}\sigma_{i+1}^{z} +
\mathcal{O}(\frac1{\Delta})
\ee
contributes to the eigenvalues with $0$ for each pair of parallel spins and with
$+1$ for each pair of antiparallel spins; these values are obtained using
the classical model (\ref{ising}). In other words, antiparallel spins increase
the energy. This means that the ground state is one of the two saturated ferromagnets
\begin{gather}
\mid \uparrow \uparrow \ldots \uparrow \rangle=\binom{1}{0}\otimes\binom{1}{0}
\otimes\ldots \otimes \binom{1}{0}\\[3mm]
\mid \downarrow \downarrow \ldots \downarrow \rangle=\binom{0}{1}\otimes\binom{0}{1}
\otimes\ldots \otimes \binom{0}{1}
\end{gather}

If $\Delta\ll -1$, the situation is the opposite and adjacent antiparallel spins
are favored so the ground state will be an anti-ferromagnet. Its actual form
is not necessarily trivial, as shown in Exercice 9;
moreover it depends on $\Delta$ and on the parity of the chain length $N$.
First, we consider the Ising limit $\Delta\rightarrow -\infty$ with $N$ even.
In that case, the ground state is a succession of up and down spins
\begin{gather}\label{antif}
\mid \uparrow \downarrow \uparrow \downarrow \ldots \rangle\\[3mm]
\mid \downarrow \uparrow \downarrow \uparrow \ldots \rangle.\nonumber
\end{gather}
Theses states are called N\'eel states.
For $N$ being even, both vectors are compatible with the periodic
boundary conditions. If $N$ is odd and if the site 1 has spin up, the site $N$
has also spin up, however sites 1 and $N$ are adjacent therefore two up spins meet.
The perfect alternation of up and down cannot be realized.
This phenomenon is called frustration (see also Exercise 8).

When $\Delta$ is finite, the situation is even more complicated:
the vectors (\ref{antif})
are not eigenvectors of the Hamiltonian so they cannot be the ground state.
The two-sites example is clear: in (\ref{autovalori}) the state
$\mid\uparrow\downarrow\rangle$ is not eigenvector. It becomes eigenvector
only at the limit, because the two eigenvalues $\Delta\pm 2$ become
degenerate so one can take linear combinations of the corresponding
eigenvectors.



\section{Yang-Baxter equation and the braid group}

\subsection{Yang-Baxter equation}

We wish now to introduce a more abstract and general formalism, that
will later allow us to investigate the spectrum of the Hamiltonian (\ref{YB1})
and its integrable properties.
Within this formalism, we make use of a special matrix, usually indicated as
$R$-matrix, and satisfies the Yang-Baxter equation \cite{baxter}.
This equation provides a set of very strong conditions on the model, implying its integrability.
This section will be developed independently of the ideas introduced in the
previous sections. The connection between the two formalisms will be done later.

Let us now introduce the fundamental relation in our context (quantum inverse scattering method), that is
the Yang-Baxter equation (YBE) \cite{korepin, baxter}
\be
R_{12}(\l_1-\l_2)\ R_{13}(\l_1-\l_3)\ R_{23}(\l_2-\l_3)=R_{23}(\l_2-\l_3)\ R_{13}(\l_1-\l_3)\ R_{12}(\l_1-\l_2),
\ee
\no where $R(\l)$ is a matrix acting on $V \tp V$. YBE acts on $V\otimes V\otimes V$, and according to the notation
introduced earlier $R_{12} = R \otimes {\mathbb I}$, $~R_{23} = {\mathbb I }\otimes R$ and so on.
We set henceforth $\l_3=0$ for simplicity.

Graphically,
one represents $R_{12}$ as

\vspace{1cm}

\begin{center}
\begin{picture}(1,1)
	\put(-120,0){\line(1,0){80}} \put(-80,40){\line(0,-1){80}}
	\put(-122,7){$1$} \put(-92,-37){$2$}
	\put(-50,30){$$}
\end{picture}
\end{center}

\vspace{2cm}

\no The YBE is then simply represented as

\vspace{1cm}

\begin{center}
\begin{picture}(1,1)
	\put(-120,40){\line(1,-1){80}} \put(-120,-40){\line(1,1){80}}
	\put(-100,40){\line(0,-1){85}}
	\put(-122,25){$2$} \put(-125,-36){$1$} \put(-95,-42){$3$}
	\put(-15,2){\line(1,0){10}} \put(-15,-2){\line(1,0){10}}
	\put(20,40){\line(1,-1){80}} \put(20,-40){\line(1,1){80}}
	\put(80,40){\line(0,-1){85}}
	\put(18,25){$2$} \put(15,-36){$1$} \put(70,-42){$3$}
\end{picture}
\end{center}

\vspace{2cm}

From the physical viewpoint, as pointed out in \cite{zamo} the Yang-Baxter equation
describes the
factorization of multi-particle scattering, a unique feature displayed by 2-d integrable systems (see figure above).
On the other hand from a mathematical viewpoint the algebraic structures underlying the Yang-Baxter
equation may be seen as deformations of the usual Lie algebras or their infinite dimensional extensions,
the Kac-Moody algebras \cite{kac}. Such deformed algebraic structures are endowed with a non trivial co-product
as we shall see, and are known as quantum groups or quantum algebras \cite{jimbo, drinf}.

\paragraph{Exercise 1:} Show that \cite{yang}
\be
R(\l)=\l\mathbb{I}+i\mathcal{P}
\ee
\no is a solution to the Yang-Baxter equation (recall ${\cal P}$ is the permutation operator).

The YBE can be also written in an alternative form. First, define
$\check{R}=\mathcal{P}R$ \cite{jimbo};
substituting in the YBE and exploiting the properties of $\mathcal{P}$, one
finds that $\check{R}$ satisfies
\be
\check{R}_{12}(d)\ \check{R}_{23}(\l_1)\ \check{R}_{12}(\l_2)=\check{R}_{23}(\l_2)\ \check{R}_{12}(\l_1)\
\check{R}_{23}(d),
\label{ybe2}
\ee
\no where $d \equiv \l_1-\l_2$. By finding solutions of the YBE in the form above (\ref{ybe2}),
then one automatically finds solutions to the original YBE. This will be the subject of the subsequent section.

\paragraph{Exercise 2:} Let
\be
R(\l) =
\begin{pmatrix} \l+i &0 &0 &0  \cr 0& \l & i  &0 \cr 0& i & \l &0  \cr 0&0 &0 & \l+i
\end{pmatrix}.
\ee
\no Show that this $R$-matrix generates the XXX spin chain Hamiltonian
\be
\left.{d\ov d\l}\check{R}_{12}(\l)\right|_{\l=0} =\mathcal{P}   \propto H_{12}\,,
\ee
up to a constant shift in (\ref{hamilxxx}) or equivalently after taking $c=0$ in (\ref{duesiti}).

The solution of the YBE is one of the primary objectives in this context, thus finding suitable methods to extract
solutions in an elegant and economical way is a significant issue. We shall discuss below how one
can identify solutions of the Yang-Baxter equations using a quite powerful technique involving the so
called braid group
(see e.g. \cite{hecke, hecke1, tl, jimbo, martinbook}).

\subsection{Braid groups}

\paragraph{Definition 3.1.} {\it The $A$-type Artin braid group is defined by generators
$g_i,\ ~ i =1, 2, \dots, N-1$,
and exchange relations:}
\ba
g_i\ g_{i+1}\ g_i & = & g_{i+1}\ g_i\ g_{i+1}, \qquad i\in \{1, \dots, N-2\},\cr
[g_i,\ g_j] & = & 0, \qquad |i-j|>1.
\ea
\no
One easily observes the structural similarity between the `braid relation' --the first of the relations above--
and the
modified YBE, which is satisfied by $\check{R}$. One can exploit this similarity, and search for
candidate solutions of
the YBE,
within the representations of the braid group.

\paragraph{Graphical representation of the braid group:}  we can graphically depict the
braid group, by defining suitable graphical representations for the
generator $g_i$ and its inverse $g^{-1}_i$. \\
\\
Depict $g_i$ as

\begin{center}
\begin{picture}(1,60)(0,10)
	\put(0,0){\line(1,0){120}} \put(0,0){\line(-1,0){120}}
	\put(0,60){\line(1,0){120}} \put(0,60){\line(-1,0){120}}
	\put(-90,60){\line(0,-1){60}} \put(90,60){\line(0,-1){60}}
	\put(-65,25){$\cdots$} \put(55,25){$\cdots$}
	\put(-90,65){$1$} \put(80,65){$N$}
	\put(-30,65){$i$} \put(15,65){$i+1$}
	\qbezier(-25,60)(-28,40)(0,30) \qbezier(0,30)(28,20)(25,0)
	\qbezier(25,60)(25,44)(3,33) \qbezier(-3,27)(-25,21))(-25,0)
\end{picture}
\end{center}

\no and $g_i^{-1}$ as

\begin{center}
\begin{picture}(1,60)(0,10)
	\put(0,0){\line(1,0){120}} \put(0,0){\line(-1,0){120}}
	\put(0,60){\line(1,0){120}} \put(0,60){\line(-1,0){120}}
	\put(-90,60){\line(0,-1){60}} \put(90,60){\line(0,-1){60}}
	\put(-65,25){$\cdots$} \put(55,25){$\cdots$}
	\put(-90,65){$1$} \put(80,65){$N$}
	\put(-30,65){$i$} \put(15,65){$i+1$}
	\qbezier(-25,60)(-26,37)(-3,33) \qbezier(3,27)(26,18)(25,0)
	\qbezier(25,60)(25,44)(0,30) \qbezier(0,30)(-25,21))(-25,0)
\end{picture}
\end{center}

\no The group identity $\mathbb{I}$ will be

\begin{center}
\begin{picture}(1,60)(0,10)
	\put(0,0){\line(1,0){120}} \put(0,0){\line(-1,0){120}}
	\put(0,60){\line(1,0){120}} \put(0,60){\line(-1,0){120}}
	\put(-90,60){\line(0,-1){60}} \put(90,60){\line(0,-1){60}}
	\put(-65,25){$\cdots$} \put(55,25){$\cdots$}
	\put(-25,60){\line(0,-1){60}} \put(25,60){\line(0,-1){60}}
	\put(-90,65){$1$} \put(80,65){$N$}
	\put(-30,65){$i$} \put(15,65){$i+1$}
\end{picture}
\end{center}

\no and two diagrams that can be brought to coincide by ``pulling the wires''
will be considered as the same group element.

Using these graphical representations, one can prove the
braid relations satisfied by the generators $g_i$. For example,
it is easily seen that $g_i\ g^{-1}_i=\mathbb{I}$


\begin{center}
\begin{picture}(1,120)(0,10)
	\put(0,0){\line(1,0){120}} \put(0,0){\line(-1,0){120}}
	\put(0,60){\line(1,0){120}} \put(0,60){\line(-1,0){120}}
	\put(0,120){\line(1,0){120}} \put(0,120){\line(-1,0){120}}
	\put(-90,60){\line(0,-1){60}} \put(90,60){\line(0,-1){60}}
	\put(-90,120){\line(0,-1){60}} \put(90,120){\line(0,-1){60}}
	\put(-65,25){$\cdots$} \put(55,25){$\cdots$}
	\put(-65,85){$\cdots$} \put(55,85){$\cdots$}
	\put(-90,125){$1$} \put(80,125){$N$} 
	\put(-30,125){$i$} \put(15,125){$i+1$} 
	\put(-90,-15){$1$} \put(80,-15){$N$} 
	\put(-30,-15){$i$} \put(15,-15){$i+1$} 
  \qbezier(-25,120)(-28,100)(0,90) \qbezier(0,90)(28,80)(25,60) 
	\qbezier(25,120)(25,104)(3,93) \qbezier(-3,87)(-25,81))(-25,60) 
	\qbezier(-25,60)(-26,37)(-3,33) \qbezier(3,27)(26,18)(25,0) 
	\qbezier(25,60)(25,44)(0,30) \qbezier(0,30)(-25,21))(-25,0) 
\end{picture}
\end{center}

\vspace{0.4cm}

\no A quick look to the diagram of $g_i^2$ explains the origin of the name braid group.

We can also show the braid relation
\be
g_i\ g_{i+1}\ g_i=g_{i+1}\ g_i\ g_{i+1},
\label{braid_g}
\ee
\no graphically.

Exploiting the graphical representation of $g_i$, the LHS becomes


\begin{center}
\begin{picture}(1,180)(0,10)
	\put(0,0){\line(1,0){120}} \put(0,0){\line(-1,0){120}}
	\put(0,60){\line(1,0){120}} \put(0,60){\line(-1,0){120}}
	\put(0,120){\line(1,0){120}} \put(0,120){\line(-1,0){120}}
	\put(0,180){\line(1,0){120}} \put(0,180){\line(-1,0){120}}
	\put(-75,180){\line(0,-1){60}} \put(75,180){\line(0,-1){60}}
	\put(-75,120){\line(0,-1){60}}
	\put(-75,60){\line(0,-1){60}} \put(75,60){\line(0,-1){60}}
	\put(-25,120){\line(0,-1){60}}
	\put(-90,185){$i-1$} \put(65,185){$i+2$} 
	\put(-30,185){$i$} \put(15,185){$i+1$} 
	\put(-90,-15){$i-1$} \put(65,-15){$i+2$} 
	\put(-30,-15){$i$} \put(15,-15){$i+1$} 
  \qbezier(-25,180)(-28,160)(0,150) \qbezier(0,150)(28,140)(25,120) 
	\qbezier(25,180)(25,164)(3,153) \qbezier(-3,147)(-25,141)(-25,120) 
	\qbezier(25,120)(28,100)(50,90) \qbezier(50,90)(78,80)(75,60) 
	\qbezier(75,120)(75,104)(53,93) \qbezier(47,87)(25,81)(25,60) 
	\qbezier(-25,60)(-28,40)(0,30) \qbezier(0,30)(28,20)(25,0)
	\qbezier(25,60)(25,44)(3,33) \qbezier(-3,27)(-25,21)(-25,0)
\end{picture}
\end{center}


\no while the RHS is


\begin{center}
\begin{picture}(1,180)(0,10)
	\put(0,0){\line(1,0){120}} \put(0,0){\line(-1,0){120}}
	\put(0,60){\line(1,0){120}} \put(0,60){\line(-1,0){120}}
	\put(0,120){\line(1,0){120}} \put(0,120){\line(-1,0){120}}
	\put(0,180){\line(1,0){120}} \put(0,180){\line(-1,0){120}}
	\put(-75,180){\line(0,-1){60}} \put(-25,180){\line(0,-1){60}}
	\put(-75,120){\line(0,-1){60}}
	\put(-75,60){\line(0,-1){60}} \put(-25,60){\line(0,-1){60}}
	\put(75,120){\line(0,-1){60}}
	\put(-90,185){$i-1$} \put(65,185){$i+2$} 
	\put(-30,185){$i$} \put(15,185){$i+1$} 
	\put(-90,-15){$i-1$} \put(65,-15){$i+2$} 
	\put(-30,-15){$i$} \put(15,-15){$i+1$} 
  \qbezier(25,180)(22,160)(50,150) \qbezier(50,150)(78,140)(75,120) 
	\qbezier(75,180)(75,164)(53,153) \qbezier(47,147)(25,141)(25,120) 
	\qbezier(-25,120)(-22,100)(0,90) \qbezier(0,90)(28,80)(25,60) 
	\qbezier(25,120)(25,104)(3,93) \qbezier(-3,87)(-25,81)(-25,60) 
	\qbezier(25,60)(22,40)(50,30) \qbezier(50,30)(78,20)(75,0)
	\qbezier(75,60)(75,44)(53,33) \qbezier(47,27)(25,21)(25,0)
\end{picture}
\end{center}

\vspace{0.4cm}

\no It is then straightforward to
check the equality of the two sides by comparing the two graphical representations.
We mentioned earlier in the text that representations of the braid group may provide solutions of the YBE.
However, the braid group
is too `big' to be physical, hence we shall restrict ourselves to quotients of the braid group to search for
solutions of the YBE \cite{hecke, hecke1, tl}.

\paragraph{Definition 3.2.} {\it The $A$-type Hecke algebra $H_{N}(q)$ is defined by the generators $g_i$,
$~i \in \{1, \ldots, N-1\}$ and
the braid relations presented above, plus an extra condition}
\ba
&& g_i\ g_{i+1}\ g_i  = g_{i+1}\ g_i\ g_{i+1}, \qquad i\in \{1, \dots, N-2\},\cr
&& (g_i-q)(g_i+q^{-1}) = 0,\cr
&& [g_i,\ g_j]  =  0, \qquad |i-j|>1.
\ea
It is clear that the Hecke algebra $H_{N}(q)$ is a quotient of the $A$-type braid group.

\no There is an alternative form of the Hecke algebra. Renaming the generators
as $U_i = g_i - q$, we get
\ba
U_i\ U_{i+1}\ U_i-U_i & = & U_{i+1}\ U_i\ U_{i+1}-U_{i+1}\cr
U_i^2 & = & -(q+q^{-1})U_i\cr
[U_i,\ U_j] & = & 0, \qquad |i-j|>0. \label{basic}
\ea You may check this as an exercise.

\paragraph{Definition 3.3.} {\it The Temperley-Lieb algebra $T_N(q)$ is a quotient of the
Hecke algebra, and is defined by (\ref{basic}) and the additional requirement:}
\be
U_{i\pm1}\ U_i\ U_{i\pm1} =U_{i\pm1}.
\ee

\paragraph{Exercise 3:} The $R$-matrix for the XXZ spin chain is the following
\be
R(\lambda)=
\begin{pmatrix}
\sinh(\l+i\m) &0 &0 &0 \cr
0&  \sinh\l & e^\l\sinh i\m
&0 \cr 0& e^{-\l}\sinh i \m  & \sinh\l
&0 \cr 0&0 &0 & \sinh(\l+i\m)
\end{pmatrix}. \label{xxz}
\ee
\no Find $\check{R}$ and show that it can be written in a form (up to an irrelevant overall factor)
\be
\check{R}=e^\l g-e^{-\l}g^{-1}.
\ee
\no Show that if $U=g-q$ satisfies the Temperley-Lieb algebra, then $\check{R}$ satisfies
the Yang-Baxter equation.

Notice that the $R$ matrix (\ref{xxz}) is expressed in the so-called homogeneous gradation,
there is also the principal gradation.
The two are related via a simple gauge transformation as\footnote{In general we have:
\be
R_{12}^{(p)}(\lambda-\mu) = V_1(-\lambda)\ V_2(-\mu)\ R_{12}^{(h)}(\lambda)\ V_1(\lambda)\ V_2(\mu)
\ee}
\be
R_{12}^{(p)}(\lambda) = V_1(-\lambda)\ R_{12}^{(h)}(\lambda)\ V_1(\lambda), ~~~~V(\lambda)= \mbox{diag}
(e^{\lambda \over 2},\
e^{-{\lambda \over2 }}).
\ee

\paragraph{Baxterization:}
For any representation $g_i$ of the $A$-type Hecke algebra we obtain a solution of the YBE,
 expressed as \cite{jimbo}:
\be
\check R_{i i+1} = e^{\lambda} g_i - e^{-\lambda} g_i^{-1}.
\ee

\paragraph{Exercise 4:} Suppose that the $R$-matrix has the form
\be
\check R_{i,i+1}(\l)=a(\l)\mathbb{I}+b(\l)U_i, \label{tl0}
\ee
\no where $U_i\in H_N(q)$. Find $a(\l),b(\l)$ so that $\check R_{i,i+1}(\l)$ satisfies
the Yang-Baxter equation.

\paragraph{Exercise 5:} Show that
\ba
\left.{d\ov d\l}\check{R}_{12}^{XXZ}(\l)\right|_{\l=0} & \propto &  H_{12}^{XXZ}\cr
 & = & {1\ov2}(\s_1^x\s_2^x+\s_1^y\s_2^y+\cosh(i\m ) \s_1^z\s_2^z + c_1(\s_1^z-\s_2^z)+c_2),
\ea
\no that is, $\check{R}^{XXZ}$ generates the XXZ spin chain Hamiltonian
(up to an additive constant). Determine
the constants $c_1$ and $c_2$.

\paragraph{Graphical Representation of Temperley-Lieb algebra:}
just as in the case of the Hecke algebra,
there is a nice graphical representation of the Temperley-Lieb algebra. The
generator $U_i$ is graphically depicted as:


\begin{center}
\begin{picture}(1,60)(0,10)
	\put(0,0){\line(1,0){120}} \put(0,0){\line(-1,0){120}}
	\put(0,60){\line(1,0){120}} \put(0,60){\line(-1,0){120}}
	\put(-90,60){\line(0,-1){60}} \put(90,60){\line(0,-1){60}}
	\put(-65,25){$\cdots$} \put(55,25){$\cdots$}
	\put(-90,65){$1$} \put(80,65){$N$}
	\put(-30,65){$i$} \put(15,65){$i+1$}
	\put(0,60){\oval(60,40)[b]}
	\put(0,0){\oval(60,40)[t]}

\end{picture}
\end{center}

The relation $U_i^2=-(q-q^{-1}) U_i$ is represented as


\begin{center}
\begin{picture}(1,120)(0,10)
	\put(0,0){\line(-1,0){200}}
	\put(0,60){\line(-1,0){200}}
	\put(0,120){\line(-1,0){200}}
	\put(-190,120){\line(0,-1){120}} \put(-10,120){\line(0,-1){120}}
	\put(-55,25){$\cdots$} \put(-155,25){$\cdots$}
	\put(-55,85){$\cdots$} \put(-155,85){$\cdots$}
	\put(-190,125){$1$} \put(-20,125){$N$}
	\put(-130,125){$i$} \put(-85,125){$i+1$}
	\put(-100,60){\oval(60,40)[b]} 
	\put(-100,0){\oval(60,40)[t]} 
	\put(-100,120){\oval(60,40)[b]} 
	\put(-100,60){\oval(60,40)[t]} 
	\put(8,62){\line(1,0){10}} \put(8,58){\line(1,0){10}}
	\put(50,60){\circle{40}}
	\put(90,100){\line(0,-1){80}} \put(170,100){\line(0,-1){80}}
	\put(130,100){\oval(40,30)[b]} \put(130,20){\oval(40,30)[t]}
	\put(90,100){\line(1,0){80}} \put(90,20){\line(1,0){80}}
\end{picture}
\end{center}
The circle in the RHS of the graph above represents the constant $~-(q+q^{-1})$.

\paragraph{Exercise 6:} Prove the relation
\be
U_i\ U_{i+1}\ U_i=U_i
\ee
\no by using the graphical representation of $U_i$.

\paragraph{Exercise 7:}
Consider the $n\times n$ matrix:
\be
{\mathrm U} = \sum_{i \neq j=1}^n(e_{ij} \otimes e_{ji} - q^{-sgn(i-j)} e_{ii} \otimes e_{jj}).
\ee
Show that it provides a representation of the Hecke algebra,
$\pi: H_N(q) \hookrightarrow \mbox{End}(({\mathbb C}^n)^{\otimes N})$
\be
\pi(U_i)= {\mathbb I} \otimes \ldots \otimes {\mathbb I} \otimes \underbrace{{\mathrm U}}_{i,\ i+1}
\otimes \ldots \otimes {\mathbb I}.
\ee

\section{Quantum integrability}

\subsection{The quantum Lax operator}

It is easy to verify that the XXX $R$-matrix may be expressed in terms of the spin ${1 \over 2}$ representation
of $\mathfrak{sl}_2$. This observation gives us a motivation to
introduce objects that are associated to higher representations of $\mathfrak{sl}_2$.
Of course such generalizations occur for any
Lie algebra,
but we use here the $\mathfrak{sl}_2$ algebra as a pedagogical example.

Take the $R$-matrix of the XXX spin chain \cite{yang},
\be
R=\l\mathbb{I}+i\mathcal{P}.
\ee
\no The permutation operator may be also expressed as:
\be
\mathcal{P}=
\begin{pmatrix}
1 &0 &0 &0 \cr 0&0 & 1 &0 \cr 0& 1 &0 &0 \cr 0&0 &0 & 1
\end{pmatrix}
=
\begin{pmatrix}
{\s^z\ov2}+{1\ov2} & \s^- \cr \s^+ & -{\s^z\ov2}+{1\ov2}
\end{pmatrix}.
\ee
\no Inspired by the above form, we introduce a
general matrix $\mathbb{P}$ as
\be
\mathbb{P}=
\begin{pmatrix}
J^z+{1\ov2} & J^- \cr J^+ & -J^z+{1\ov2}
\end{pmatrix}, \label{matrixp}
\ee
\no where $J^{\pm}, J^z$ are now abstract algebraic elements, which satisfy as will be clear below the
$\mathfrak{sl}_2$ exchange relations (\ref{rels}).
Now define the Lax operator $L$ as
\be
L=\l\mathbb{I}+i\mathbb{P}, \label{lmatrix0}
\ee
\no and assume $L$ satisfies the following fundamental algebraic relation \cite{FT, FTS, tak}
\be
R_{12}(\l_1-\l_2)\ L_{1n}(\l_1)\ L_{2n}(\l_2)=L_{2n}(\l_2)\ L_{1n}(\l_1)\ R_{12}(\l_1-\l_2).
\label{RLL}
\ee
In (\ref{RLL}) the indices 1, 2 traditionally denote the auxiliary space, and $n$ denotes
the quantum space on which the algebra
generators act. In general $L \in \mbox{End}(V) \otimes {\cal A}$, where ${\cal A}$
is the algebra defined by (\ref{RLL}). Different choices of $R$ matrix lead naturally to distinct algebras,
as will be transparent later in the text.

Graphically,
one represents the $L$ operator as

\vspace{1cm}

\begin{center}
\begin{picture}(1,1)
	\put(-120,0){\line(1,0){80}} \multiput(-80,40)(0,-9){9}{\line(0,-1){6}}
	\put(-122,7){$1$} \put(-92,-37){$n$}
\end{picture}
\end{center}

\vspace{2cm}

\no The fundamental algebraic relation satisfied by $L$ is then simply represented graphically as

\vspace{1cm}

\begin{center}
\begin{picture}(1,1)
	\put(-120,40){\line(1,-1){80}} \put(-120,-40){\line(1,1){80}}
	\multiput(-100,40)(0,-9){10}{\line(0,-1){6}}
	\put(-122,25){$2$} \put(-125,-36){$1$} \put(-95,-42){$n$}
	\put(-15,2){\line(1,0){10}} \put(-15,-2){\line(1,0){10}}
	\put(20,40){\line(1,-1){80}} \put(20,-40){\line(1,1){80}}
	\multiput(80,40)(0,-9){10}{\line(0,-1){6}}
	\put(18,25){$2$} \put(15,-36){$1$} \put(70,-42){$n$}
\end{picture}
\end{center}

\vspace{2cm}

\no For the particular choice of $R$ matrix the YBE lives in:
$\mathbb{C}^2\tp \mathbb{C}^2\tp \mathcal{Y}$,
\no with $\mathcal{Y}$ being the Yangian of  $\mathfrak{sl}_2$ defined by (\ref{RLL}). The
relation above holds for
any $R$ matrix associated to any Lie algebra, but we focus here in $\mathfrak{sl}_2$ for simplicity.

Substituting the explicit forms
of $R$ and $L$ in (\ref{RLL}), one finds that the following condition should hold
\be
[\mathbb{P}_1,\ \mathbb{P}_2]=(\mathbb{P}_2-\mathbb{P}_1)\mathcal{P}_{12}=\mathcal{P}_{12}(\mathbb{P}_1-\mathbb{P}_2).
\label{Pcommut}
\ee
\no
(here the quantum space for $\mathbb{P}$ is omitted).
It may be shown that the condition above leads to the $\mathfrak{sl}_2$ exchange relations.

\paragraph{Exercise 1:} Consider $\mathbb{P}$ defined in (\ref{matrixp});
based on relations (\ref{Pcommut}) and notation (\ref{notation}) show that $J^z,\ J^{\pm}$
satisfy the $\mathfrak{sl}_2$ exchange relations.

\subsection{The q-deformed case: $U_q(\mathfrak{sl}_2)$}

We come now to the $q$-deformed case, which corresponds to the XXZ spin chain and its generalizations.
Take the $R$-matrix of the XXZ model (\ref{xxz}), which can be also written as (homogeneous gradation)
\be
R(\l) = e^\l R^+ -e^{-\l}R^- ,
\ee
\no where $R^+,\ R^-$ are upper, lower triangular matrices:
\be
R^+ =
\begin{pmatrix}
q &0 &0 &0 \cr 0& 1 & q-q^{-1} &0 \cr 0& 0 & 1 &0 \cr 0&0 &0 & q
\end{pmatrix},\qquad
R^- =
\begin{pmatrix}
q^{-1} &0 &0 &0 \cr 0& 1 & 0 &0 \cr 0& -(q-q^{-1}) & 1 &0 \cr 0& 0&0 & q^{-1}
\end{pmatrix}.
\ee
\no These matrices may be also written as
\be
R^+ =
\begin{pmatrix}
q^{{1 \ov 2}(\s^z+1)} & (q-q^{-1})\s^- \cr 0 & q^{{1\ov2}(-\s^z+1)}
\end{pmatrix},
\qquad R^-=
\begin{pmatrix}
q^{-{1\ov2}(\s^z-1)} & 0 \cr -(q-q^{-1})\s^+ & q^{{1\ov2}(\s^z-1)}
\end{pmatrix}.
\ee

Now suppose that the $L$-matrix can be also written in terms of upper/lower triangular matrices, giving
rise as will be clear to upper/lower Borel subalgebras of $U_q(\mathfrak{sl}_2)$ \cite{FTS, tak},

\be
L=e^\l L^+-e^{-\l}L^-,
\ee
\no where
\be
L^+=
\begin{pmatrix}
c A & B \cr 0 & c D
\end{pmatrix},
\qquad
L^-=
\begin{pmatrix}
c^{-1} A^{-1} & 0 \cr C & c^{-1} D^{-1}
\end{pmatrix}.
\ee
where $c$ is an arbitrary constant.
\no
These matrices $R,L$ obey (\ref{RLL}).
By taking various limits of it  when $\l_i \to \pm \infty$, one arrives at the following set of
equations:
\begin{itemize}
	\item Limit $\l_1- \l_2,\l_1,\l_2\to\infty$
	\be
	R_{12}^+\ L_{1n}^+\ L_{2n}^+=L_{2n}^+\ L_{1n}^+\ R_{12}^+, \label{11}
	\ee
	\item Limit $\l_1,\l_2\to-\infty$ and $\l_1-\l_2\to-\infty$
	\be
	R_{12}^-\ L_{1n}^-\ L_{2n}^-=L_{2n}^-\ L_{1n}^-\ R_{12}^-, \label{12}
	\ee
	\item Limit $\l_1\to\infty,\l_2\to-\infty$
	\be
	R_{12}^+\ L_{1n}^+\ L_{2n}^-=L_{2n}^-\ L_{1n}^+\ R_{12}^+.\label{13}
	\ee
\end{itemize}
Note that the Lax operator can be also expressed in the principal gradation via:
\be
L_{1n}^{(p)}(\lambda) =  V_1(-\lambda)\ L^{(h)}_{1n}(\lambda)\ V_1(\lambda). \label{lgauge}
\ee

\paragraph{Exercise 2:} Solve the above set of equations (\ref{11})-(\ref{12}) and determine the various relations
among $A,\ B,\ C$ and $D$:
\be
[B,\ C]=(q-q^{-1})(A^2-D^2), \qquad A\ D= D\ A ={\mathbb I}, \qquad A\ B=qB\ A, \qquad A\ C = q^{- 1} C\ A.
\ee
\no By further imposing
\be
A=q^{J^z}=D^{-1}, \qquad B=(q-q^{-1}) J^-,\qquad C=(q-q^{-1}) J^+,
\ee
\no these are in fact the relations of the deformed Lie group $U_q(SL_2)$. Show that the latter
relations imply also:
\ba
[J^+,\ J^-]  &=& {q^{2J^z}-q^{-2J^z}\ov q-q^{-1}}\cr
[J^z,\ J^{\pm}] &=& \pm J^{\pm}.
\ea
\no This is the so-called q-deformed $\mathfrak{sl}_2$ algebra denoted as $U_q(\mathfrak{sl}_2)$ \cite{jimbo}.
Finally, taking the limit $q\to 1$ one recovers the familiar $\mathfrak{sl}_2$ relations.
$U_q(\mathfrak{sl}_2)$ is a Hopf algebra and is equipped with a non-trivial co-product \cite{jimbo},
$\Delta: U_q(\mathfrak{sl}_2) \hookrightarrow  U_q(\mathfrak{sl}_2) \otimes  U_q(\mathfrak{sl}_2)$ such that
\ba
&& \Delta(q^{\pm J^z}) = q^{\pm J^z} \otimes q^{\pm J^z}, \cr
&& \Delta(J^{\pm}) = q^{-J^z} \otimes J^{\pm} + J^{\pm} \otimes q^{J^z}. \label{copr}
\ea
More details on co-products of quantum algebras will be given in subsequent lecture.

\paragraph{Exercise 3:} Show that the quantity
\be
C= q q^{2J^z} + q^{-1}q^{-2J^z} + (q-q^{-1})^2 J^- J^+.
\ee
is the Casimir operator of $U_q(\mathfrak{sl}_2)$

\paragraph{Exercise 4:} Show that if $J^z,\ J^{\pm}$ satisfy the $U_q(\mathfrak{sl}_2)$ algebra
then $\Delta(J^z),\ \Delta(J^{\pm})$ also satisfy $U_q(\mathfrak{sl}_2)$.

\subsection{The transfer matrix: integrability}

Our purpose now is to construct and solve 1-dimensional spin chain-like systems using the
so-called quantum inverse scattering method (see e.g. \cite{korepin, tak, faddeev1, faddeev2}).
To achieve this we shall introduce tensor
type representations of the fundamental algebraic relation (\ref{RLL}).

We may introduce the so-called \emph{monodromy} matrix, as
\be
T_a(\l)=L_{aN}(\l)\ L_{aN-1}(\l)\cdots L_{a1}(\l).
\ee
\no
and apparently $T(\lambda) \in \mbox{End}(V) \otimes {\cal A}^{\otimes N}$.
As customary we have suppressed the quantum spaces $1, \ldots, N$ from the monodromy matrix.
The monodromy matrix can be graphically represented as

\vspace{1cm}

\begin{center}
\begin{picture}(1,1)
	\put(-90,0){\line(1,0){180}}
	\multiput(-75,30)(0,-8){8}{\line(0,-1){6}}
	\multiput(-45,30)(0,-8){8}{\line(0,-1){6}}
	\multiput(75,30)(0,-8){8}{\line(0,-1){6}}
	\put(-15,-20){$\cdots\cdots\cdots\cdots$}
	\put(-78,-50){$N$}
	\put(-48,-50){$N-1$}
	\put(72,-50){$1$}
	\put(-92,5){$a$}
\end{picture}
\end{center}

\vspace{1.5cm}

\no and also satisfies the fundamental algebraic relation (FRT)
\be
R_{12}(\l_1-\l_2)\ T_1(\l_1)\ T_2(\l_2)=T_2(\l_2)\ T_1(\l_1)\ R_{12}(\l_1-\l_2).
\label{rtt}
\ee
\no Graphically, the proof is immediate, if one takes into account the
graphical representation of the $RLL=LLR$ relation, presented above.

\vspace{1cm}

\begin{center}
\begin{picture}(1,1)
	\put(-200,20){\line(1,-1){40}} \put(-200,-20){\line(1,1){40}}
	\put(-160,20){\line(1,0){130}}
	\put(-160,-20){\line(1,0){130}}
	\multiput(-155,30)(0,-8){8}{\line(0,-1){6}}
	\multiput(-130,30)(0,-8){8}{\line(0,-1){6}}
	\multiput(-40,30)(0,-8){8}{\line(0,-1){6}}
	\put(-100,-3){$\cdots\cdots$}
	\put(-160,-50){$N$}
	\put(-132,-50){$N-1$}
	\put(-42,-50){$1$}
	\put(-208,20){$a$}
	\put(-208,-20){$b$}
	\put(-10,2){\line(1,0){10}}
	\put(-10,-2){\line(1,0){10}}
	\put(17,20){\line(1,0){130}}
	\put(17,-20){\line(1,0){130}}
	\multiput(25,30)(0,-8){8}{\line(0,-1){6}}
	\multiput(50,30)(0,-8){8}{\line(0,-1){6}}
	\multiput(140,30)(0,-8){8}{\line(0,-1){6}}
	\put(80,-3){$\cdots\cdots$}
	\put(147,20){\line(1,-1){40}} \put(147,-20){\line(1,1){40}}
	\put(20,-50){$N$}
	\put(48,-50){$N-1$}
	\put(138,-50){$1$}
	\put(190,20){$b$}
	\put(190,-20){$a$}
\end{picture}
\end{center}

\vspace{1.5cm}

\no On the other hand, FRT can be also proved algebraically, consider $N=2$ for simplicity:
\ba
R_{ab}T_aT_b & = & R_{ab}L_{a2}L_{a1}L_{b2}L_{b1}=R_{ab}L_{a2}L_{b2}L_{a1}L_{b1}\cr
 & = & L_{b2}L_{a2}R_{ab}L_{a1}L_{b1} = L_{b2}L_{a2}L_{b1}L_{a1}R_{ab} \cr
 & = & L_{b2}L_{b1}L_{a2}L_{a1}R_{ab} = T_bT_aR_{ab}.
\ea
Tracing over the auxiliary space we get the \emph{transfer matrix} $t(\l)$
\be
t(\l)=\textrm{Tr}_a[T_a(\l)].
\ee
and $t(\lambda) \in {\cal A}^{\otimes N}$.
\no The transfer matrix constitutes a one-parameter family of commuting operators
\be
[t(\l),\ t(\l')]=0.
\ee
\no The proof goes as follows:
\ba
t(\l)\ t(\l') & = & (\textrm{Tr}_aT_a(\l))\ (\textrm{Tr}_bT_b(\l'))=\textrm{Tr}_{ab}[T_a(\l)\ T_b(\l')]\cr
 & = & \textrm{Tr}_{ab}[R_{ab}(-\d)\ R_{ab}(\d)\ T_a(\l)\ T_b(\l')]\cr
 & = & \textrm{Tr}_{ab}[R_{ba}(-\d)\ T_b(\l')\ T_a(\l)\ R_{ab}(\d)]\cr
 & = & \textrm{Tr}_{ab}[T_b(\l')\ T_a(\l)]\cr
 & = & (\textrm{Tr}_{b}T_b(\l'))\ (\textrm{Tr}_{a}T_a(\l))\cr
 & = & t(\l')\ t(\l). \label{commut}
\ea
\no This condition ensures that the system at hand is integrable.

The commutation relation (\ref{commut}) holds $\forall \lambda,\lambda' \in \mathbb{C}$, which implies that
factors of formal series expansion commute with each other. Explicitly we have:\\
\begin{equation}
\label{YB68}
\sum_{n,m}\lambda^{n} \lambda^{'m}\left [t^{(n)},\ t^{(m)}\right ] = 0
\end{equation}\\
and this automatically yields:\\
\begin{equation}
\label{YB69}
\left [t^{(n)},\ t^{(m)}\right ] = 0 \qquad \forall n,\ m.
\end{equation}\\
It is clear that the elements $t^{(n)}$ are the so called charges in involution.
Expansions based on other points are also possible.
Below we shall derive the first two charges, i.e. the
momentum and energy.

\subsection{The momentum and the Hamiltonian}

If we restrict our attention to the case where both auxiliary and quantum spaces are represented to
the same vector space $V$, $L \hookrightarrow R$, then we obtain a
local Hamiltonian as long as $R(0) \propto {\cal P}$.
For instance in the $U_q(\mathfrak{sl}_2)$ case,
when both auxiliary and quantum space correspond to the fundamental representation of the algebra, we deal with the
familiar XXZ $R$-matrix (\ref{xxz}) (we shall see a particular example below).
Recall also, that since $R(0) \propto {\cal P}$ one may show that $t(0) \propto \Pi = \exp(-i{\mathbb P})$ where ${\mathbb P}$ is
the momentum of the system (\ref{trans}).
To conclude, the momentum and Hamiltonian
belong to the family of commuting operators obtained from the transfer matrix (see also \cite{faddeev1}).

In particular we shall show
\be
\left.{d \over d \lambda}\ln(t(\lambda))\right|_{\lambda =0} \propto H= \sum_{j=1}^N H_{j j+1},
~~~~~\mbox{where}, ~~~~~H_{j j+1} \propto \left.{d \check R_{j j+1}(\l)\over d \lambda}\right|_{\lambda =0}
\ee
imposing manifestly periodic boundary conditions: $H_{N N+1} = H_{N1}$.

Consider in general
\be
R(0) \propto {\cal P}
\ee
Let us now compute:
\begin{equation}
\label{YB73}
\left.{\displaystyle\frac{d\ln(t(\lambda))}{d\lambda}}\right|_{\lambda=0}=t^{-1}(0)t'(0)
\end{equation}
We now want to compute this expression in detail. First consider:
\ba
\label{YB74}
t(0)  & = & {\rm Tr}_0\ [R_{0N}(0)\ R_{0\ N-1}(0) \ldots  R_{01}(0)] \cr
 & \propto & {\rm Tr}_{0}\ [{\cal P}_{0N}\ {\cal P}_{0\ N-1} \ldots {\cal P}_{01}]  \cr
& = & {\rm Tr}_{0}\ [{\cal P}_{N\ N-1}\ {\cal P}_{N\ N-2} \dots {\cal P}_{N1}\ {\cal P}_{0N}]  \cr
 & = & {\cal P}_{N\ N-1}\ {\cal P}_{N\ N-2} \dots {\cal P}_{N1}\ Tr_{0}{\cal P}_{0N}  \cr
&=&  {\cal P}_{12}\ {\cal P}_{23} \ldots {\cal P}_{N-1\ N}= \Pi
\ea
where we have used the fact that ${\rm Tr}_0 {\cal P}_{0N} \propto {\mathbb I}$
for all known physical systems (for instance for
all the solutions emanating from the Hecke algebras). Recall that $\Pi = e^{-i{\mathbb P}}$ is
the translation operator
and ${\mathbb P}$ the momentum of the system.
As a straightforward consequence we have:
\ba
\label{YB75}
t^{-1}(0) & \propto & {\cal P}_{1N}\ {\cal  P}_{1\;N-1} \dots {\cal P}_{13}\ {\cal P}_{12}
\cr & = & {\cal P}_{N-1 N}\ {\cal  P}_{N-2\;N-1} \dots {\cal P}_{23}\ {\cal P}_{12}.
\ea
Now we have to compute the derivative of $t(\lambda)$, namely:
\ba
\label{YB76}
t'(0) & = & {\rm Tr}_{0}\sum_{i} R_{0N}(0) \ldots R'_{0i}(0) \ldots R_{01}(0) = \cr
& \propto & {\rm Tr}_{0} \sum_{i} {\cal P}_{0N} \ldots R'_{0i}(0) \ldots {\cal P}_{01} \cr
& = & {\rm Tr}_{0} \sum_{i} {\cal P}_{N\ N-1} \ldots {\cal P}_{N\;i+1}\
R'_{Ni}(0) \ldots {\cal P}_{N1}\ {\cal P}_{0N} = \cr
& = & \sum_{i}  {\cal P}_{N\ N-1} \ldots {\cal P}_{N\;i+1}\
R'_{Ni}(0) \ldots {\cal P}_{N1}.
\ea

Collecting all previous results we can write down:
\ba
\label{YB77}
t^{-1}(0)\ t'(0)  & \propto & \sum_{i} {\cal P}_{N 1 }\ {\cal P}_{N 2}\ldots {\cal P}_{N\ N-2}\
{\cal P}_{N N-1}\ {\cal P}_{N\ N-1} \ldots
\ {\cal P}_{N\ i+1} \ldots R'_{Ni} \ldots {\cal P}_{N1}= \cr
& = & \ldots \cr
& = & \sum_{i }\check R'_{i\ i+1}(0).
\ea

We shall provide here as an example the Hamiltonian of the XXZ model.
Let us recall the $R$-matrix of the XXZ chain (principal gradation):
\begin{equation}
\label{YB70}
\begin{array}{l}
R(\lambda) =
\begin{pmatrix}
\sinh(\lambda+i\mu) & 0 & 0 & 0 \\
0 & \sinh \lambda & \sinh i\mu & 0 \\
0 & \sinh i \mu & \sinh \lambda & 0 \\
0 & 0 & 0 & \sinh (\lambda+i\mu)\\
\end{pmatrix}
\end{array}
\end{equation}
Its derivative  at the origin is
\begin{equation}
\label{YB72}
R'(0) =
\begin{pmatrix}
\cosh i\mu& 0 & 0 & 0 \\
0 & 1 & 0 & 0 \\
0 & 0 & 1 & 0 \\
0 & 0 & 0 & \cosh i\mu\\
\end{pmatrix}
\end{equation}
and also
\begin{equation}
R(0)=\sinh i\mu\: \mathcal{P}\,,\qquad R^{-1}(0)=(\sinh i\mu)^{-1} \mathcal{P}
\end{equation}
therefore, recalling the expression of ${\cal P}$ in (\ref{perm}), we get
\begin{equation}
\begin{array}{c}
R^{-1}(0)\ R'(0)  =\frac{1}{\sinh i\mu}
\begin{pmatrix}
\cosh i\mu & 0 & 0 & 0 \\
0 & 0 & 1 & 0 \\
0 & 1 & 0 & 0 \\
0 & 0 & 0 & \cosh i\mu \\
\end{pmatrix} \\
\label{YB80}
\propto {1\over 2} \Big (\sigma^{x}\otimes\sigma^{x}+\sigma^{y}\otimes\sigma^{y} +
\cosh i\mu (\mathbb{I}\otimes\mathbb{I}+\sigma^{z}\otimes\sigma^{z}) \big ).
\end{array}
\end{equation}\\
Thus equation (\ref{YB73}) becomes:
\ba
\label{81}
&& H \propto \left.\frac{d \ln(t(\lambda))} {d\lambda}\right|_{\lambda=0} \cr
&\propto & -{1\over 2} \sum_{i} \left[\sigma_{i}^{x}\ \sigma_{i+1}^{x}+
\sigma_{i}^{y}\ \sigma_{i+1}^{y}+\cosh i \mu(\sigma_{i}^{z}\ \sigma_{i+1}^{z}+\mathbb{I}_{i}
\ \mathbb{I}_{i+1}) \right ].
\ea
This last expression corresponds to our XXZ Hamiltonian (\ref{YB1}) (up to a constant),
by fixing $\Delta = \cosh i \mu$.
Such an observation is extremely important because thanks to
(\ref{YB69}) we can construct a series of conserved quantities, simply by
computing the following derivatives:
\begin{equation}
\label{82}
\left.{\displaystyle\frac{d^{(n)}\ln(t(\lambda))}{d^{(n)}\lambda}}\right| _{\lambda = 0} \propto {H}^{(n)}.
\end{equation}
In the next section we will study the form and the properties of the ground state of the XXZ model,
by introducing the Bethe Ansatz technique.

\paragraph{Exercise 5:} Show that $[\Pi,\ H]=0$.

\section{Review on quantum algebras}

\subsection{Quantum algebras and non-trivial co-products}

It will be instructive for what follows to examine two basic classes of deformed
algebras arising in the context of integrable systems that is the so called
Yangian and the $q$-deformed algebras. It is worth mentioning that the deformed algebras
underlying any integrable system play an essential role in the context of algebraic Bethe ansatz
for finding the associated spectra,
as will be transparent in the following. Also, linear intertwining relations between the $R$ matrix and co-products
of the algebra elements can be used for the derivation of $R$ matrices --we shall briefly discuss this issue later in the text.
We shall focus here on these algebraic structures, and exhibit how their
non-trivial co-products emerge naturally in the context of quantum integrability.

\subsubsection{The Yangian}

We shall first consider the Yangian
(for a review on Yangians see e.g. \cite{drinf, molev, bele} ). The $\mathfrak{gl}_n$ Yangian ${\cal Y}$,
is a non abelian algebra with generators ${\mathrm Q}_{ab}^{(p)}$
and defining relations given below
\ba
&&\Big [ {\mathrm
Q}_{ab}^{(0)},\ {\mathrm Q}_{cd}^{(0)} \Big ] =i\delta_{cb}{\mathrm Q}^{(0)}_{ad} - i\delta_{ad}{\mathrm Q}^{(0)}_{cb}
\cr && \Big [ {\mathrm Q}_{ab}^{(0)},\ {\mathrm Q}_{cd}^{(1)} \Big ] =i\delta_{cb}
{\mathrm Q}^{(1)}_{ad} - i\delta_{ad}{\mathrm Q}^{(1)}_{cb}
\cr
  && \Big [ {\mathrm Q}_{ab}^{(1)},\ {\mathrm Q}_{cd}^{(1)} \Big ]
=i\delta_{cb}{\mathrm Q}^{(2)}_{ad} -i \delta_{ad}{\mathrm Q}^{(2)}_{cb}
+{i h^2\over 4}{\mathrm Q}_{ad}^{(0)}(\sum_{e} {\mathrm
Q}_{ce}^{(0)}{\mathrm Q}_{eb}^{(0)})- {i h^2\over 4}(\sum_{e}{\mathrm Q}_{ae}^{(0)}{\mathrm Q}_{ed}^{(0)})
{\mathrm Q}_{cb}^{(0)}
\cr && a, b \in \{ 1,\ 2, \dots, n \}
\ea
and also relations
\ba
\Big [ {\mathrm Q}_{ab}^{(0)},\ \Big [ {\mathrm Q}_{cd}^{(1)},\ {\mathrm Q}_{ef}^{(1)} \Big ] \Big ]&-& \Big
[ {\mathrm Q}_{ab}^{(1)},\
\Big [{\mathrm Q}_{cd}^{(0)},\ {\mathrm Q}_{ef}^{(1)} \Big ] \Big ] \nonumber\\ =
{h^2\over 4}\sum_{p,q} \Big ( \Big [ {\mathrm Q}_{ab}^{(0)},\ \Big [ {\mathrm Q}_{cp}^{(0)}{\mathrm Q}_{pd}^{(0)},\
{\mathrm Q}_{eq}^{(0)}{\cal
Q}_{qf}^{(0)} \Big ] \Big ] &-&\Big [ {\mathrm Q}_{ap}^{(0)}{\mathrm Q}_{pb}^{(0)},\ \Big [ {\mathrm Q}_{cd}^{(0)},\ {\mathrm
Q}_{eq}^{(0)}{\mathrm Q}_{qf}^{(0)} \Big ] \Big ]  \Big ).
\ea

The Yangian is endowed with a co-product
$\Delta: {\cal Y} \hookrightarrow {\cal Y} \otimes {\cal Y}$ such that
\ba
\Delta({\mathrm Q}_{ab}^{(0)}) &=& {\mathrm Q}_{ab}^{(0)} \otimes {\mathbb I} + {\mathbb I} \otimes
{\mathrm Q}_{ab}^{(0)} \cr
\Delta({\mathrm Q}_{ab}^{(1)}) &=& {\mathrm Q}_{ab}^{(1)}
\otimes {\mathbb I} + {\mathbb I}\otimes {\mathrm Q}_{ab}^{(1)} + {h \over 2}
\sum _{d=1}^n({\mathrm Q}_{ad}^{(0)}\otimes
{\mathrm Q}_{db}^{(0)}-
{\mathrm Q}_{db}^{(0)}\otimes {\mathrm Q}_{ad}^{(0)}), \label{cop}
\ea
Define also the opposite co-product $\Delta':\ {\cal Y}\ \hookrightarrow\ {\cal Y}^{\otimes 2}$:
\be
\Delta' = \sigma \circ \Delta
\ee
where $\sigma$ is the `shift operator', $\sigma: a \otimes b \hookrightarrow b \otimes a$.
We may also define the $l$ co-products $\Delta^{(l)},\ \Delta^{'(l)}: {\cal
Y}\ \hookrightarrow \ {\cal Y}^{\otimes (l)}$ as
\be
\Delta^{(l)} = (\mbox{id} \otimes \Delta^{(l-1)})\Delta, \qquad \Delta^{'(l)} =
(\mbox{id} \otimes \Delta^{(l-1)})
\Delta'. \label{cop22}
\ee
and obtain the $l$ co-product.

The asymptotic behavior of the monodromy matrix $T$
as $\lambda \to \infty$ provides tensor product realizations of ${\cal Y}$.
Let us briefly review how this process works. Recall that the operators $L$ and
$T$ are treated as $n \times n$ matrices with entries being elements of ${\cal Y}$,
${\cal Y}^{\otimes N}$ respectively. $L$ in particular is given by (\ref{lmatrix0}) with
${\mathbb P}_{ab} \in \mathfrak{gl}_n$.
The monodromy matrix
$T$ as $\lambda \to \infty$ may be written as
(for simplicity we suppress the `auxiliary' space index $0$ from $T$ in the following)
\ba
T(\lambda \to \infty) \propto {\mathbb I} + \sum_{m=0}^{\infty} \lambda^{-m-1}\ t^{(m)}. \label{asy}
\ea
Exchange relations among the charges $t_{ab}^{(m)}$ (the entries of $t^{(m)}$) may be derived by virtue
of the fundamental algebraic relation (\ref{rtt}), as $\lambda_{i} \to \infty$. To extract the Yangian
generators we study the asymptotic expansion (\ref{asy}) keeping higher orders in
the ${1 \over \lambda}$ expansion.
Recalling the form of $L$
 we conclude that
\ba T(\lambda \to \infty)
\propto {\mathbb I} +{i\over \lambda}\sum_{i=1}^{N} {\mathbb P}_{0i} -
{1\over \lambda^{2}} \sum_{i > j=1}^{N}
{\mathbb P}_{0i}\ {\mathbb P}_{0j} +{\cal O}({1\over \lambda ^{3}}). \label{asyt}
\ea
Now consider the quantities below written as combinations of
$t^{(p)},~~p \in \{0,\ 1 \}$,
\ba
{\mathbb Q}^{(0)} = t^{(0)}, ~~~~~{\mathbb Q}^{(1)} = t^{(1)}
-{1\over 2} {\mathbb Q}^{(0)}\ {\mathbb Q}^{(0)}
\ea
where the form of $t^{(p)}$ is defined by (\ref{asy}), (\ref{asyt}).
Then ${\mathbb Q}^{(p)}$ may be expressed as (set here $h=-1$ in
(\ref{cop}))
\ba
{\mathbb Q}^{(0)} = i\sum_{i=1}^{N} {\mathbb P}_{0i},~~~~~
{\mathbb Q}^{(1)} = {1\over 2}\sum_{i=1}^{N}{\mathbb P}_{0i}^2 +{1\over 2}\sum_{j>i=1}^{N}
({\mathbb P}_{0i}\ {\mathbb P}_{0j} - {\mathbb P}_{0j}\ {\mathbb P}_{0i}). \label{qb}
\ea
Note that for simplicity both quantum and auxiliary indices in ${\mathbb Q}^{(p)}$ (\ref{qb}) are omitted.
The entries of the matrices ${\mathbb Q}^{(p)}$ are the non-local charges ${\mathbb Q}_{ab}^{(p)} \in
{\cal Y}^{\otimes N}$
being co-product realizations of the Yangian, i.e.
\ba
{\mathbb Q}_{ab}^{(p)} = \Delta^{(N)}({\mathrm Q}_{ab}^{(p)}) ~~~~p \in \{ 0,\
1\}, ~~~~a,b \in \{1, \ldots, n\}. \label{coco}
\ea

\subsubsection{The $U_q(\mathfrak{sl}_2)$ algebra}

We saw in the preceding lecture that by taking appropriate limits of $\l \to \pm \infty$,
 one derives the $U_q(\mathfrak{sl}_2)$
relations. The monodromy matrix can be also used to derive the co-product for
the $q$ deformed case. We shall also derive here linear intertwining relations among the $L,\ R$ matrices and
the co-products of the associated deformed algebra.

First, take the limit $\l \to \infty$, $T(\l)$ becomes
\ba
T^+ & = & L_{0N}^+L_{0N-1}^+\cdots L_{01}^+\cr
		& = & \begin{pmatrix} A_N & B_N \cr 0 & D_N \end{pmatrix}
					\begin{pmatrix} A_{N-1} & B_{N-1} \cr 0 & D_{N-1} \end{pmatrix}
		\cdots \begin{pmatrix} A_1 & B_1 \cr 0 & D_1 \end{pmatrix}.
\ea
\no Take $N=2$ for simplicity, then
\be
T^+=\begin{pmatrix} A_2 A_1 & A_2 B_1+B_2 D_1 \cr 0 & D_1D_2 \end{pmatrix}.
\ee
\no By identifying
\be
A=c q^{J^z}, \qquad B=(q -q^{-1}) J^-, \qquad D= c q^{-J^z},
\ee
\no one reads then from the elements of the monodromy matrix the co-products (\ref{copr})
(see also e.g. \cite{tak, jimbo}). If the co-product is known, one can also construct the $N$ co-product,
by iteration via (\ref{cop22}).

By using these notations and ideas, we can derive important relations
between $L,\ R$ and the co-products of the algebra elements. Start from the fundamental algebraic
relation (\ref{RLL}),
and take $\l_1\to\infty,\ \l_2\to\l$, to get
\be
R_{12}^+\ L_{1n}^+\ L_{2n}(\l)=L_{2n}(\l)\ L_{1n}^+\ R_{12}^+,
\label{RLL+}
\ee
\no where
\be
R_{12}^+ =
\begin{pmatrix} \a_2 & \b_2 \cr 0 & \d_2 \end{pmatrix},\qquad
L_{1n}^+=
\begin{pmatrix} A_n & B_n \cr 0 & D_n \end{pmatrix}.
\ee
\no Now $\a,\b,\d$ are in the fundamental representation of $U_q(\mathfrak{sl}_2)$, while $A,\ B,\ D$ are
abstract elements of the algebra, that is
\ba
\a & = &     cq^{{\s^z\ov2}}  \qquad\qquad A=cq^{J^z}\cr
\d & = &    cq^{-{\s^z\ov2}} \qquad\qquad  D = c q^{-J^z} \cr
\b & = & (q-q^{-1})\s^- \qquad\qquad B = (q-q^{-1})J^-
\ea
\no Now define the representation
$\p: U_q(\mathfrak{sl}_2) \hookrightarrow \mbox{End}(\mathbb{C}^2)$
such as
\ba
\p(J^z)  =  {\s^z\ov2}, \qquad
\p(J^{\pm})  =  \s^{\pm}.
\ea
\no and also
\ba
\p(A)  =  \a, \qquad
\p(B)  =  \b, \qquad
\p(D) =  \d.
\ea
\no
Substituting in (\ref{RLL+}), we get
\be
\begin{pmatrix} \a_2A_n & \a_2B_n+\b_2D_n \cr 0 & \d_1D_n \end{pmatrix}L_{2n}(\l)=
L_{2n}(\l) \begin{pmatrix} \a_2A_n & A_n\b_2+\d_2B_n \cr 0 & \d_2D_n \end{pmatrix}.
\ee
\no The latter suggests the $L$ `commutes' with each one of the entries of the right and left matrices.
More precisely, the `commutation' with the first entry of
the matrices reads as
\be
\a_2\ A_n\ L_{2n}(\l)=L_{2n}(\l)\ \a_2\ A_n,
\ee
\no or,
\be
(q^{{\s^z\ov2}}\tp q^{J^z})\ L_{2n}(\l)=L_{2n}(\l)\ (q^{{\s^z\ov2}}\tp q^{J^z}) \label{equ1}.
\ee
\no Since
\be
\D(q^{J^z})=q^{J^z}\tp q^{J^z}\  \Rightarrow \ (\pi\tp \textrm{id})\D(q^{J^z})=q^{{\s^z\ov2}}\tp q^{J^z}.
\ee
therefore (\ref{equ1}) may be also expressed as:
\be
(\pi\tp \textrm{id})\D(q^{J^z})\ L(\lambda) = L(\lambda)\ (\pi\tp \textrm{id})\D(q^{J^z}).
\ee
\no Reading the second entry, we have
\be
(\a_2B_n+\b_2D_n)\ L_{2n}(\l)= L_{2n}(\l)\ (A_n\b_2+\d_2B_n).
\ee
\no Substituting the explicit forms of $\a,\ \b,\ A,\ D$,
\be
[\s^-\tp q^{-J^z}+q^{{\s^z\ov2}}\tp S^-]\  L_{2n}(\l) = L_{2n}(\l)\ [\s^-\tp q^{J^z}+q^{-{\s^z\ov2}}\tp S^-].
\ee
\no It is easy to observe that the term inside the parenthesis on the RHS is
just $(\p \tp \textrm{id})\D(J^-)$,
\no while the LHS is $(\p\tp \textrm{id})\D'(J^-)$. Equation (\ref{RLL+}) then
becomes
\be
(\p\tp \textrm{id})\D'(J^-)\ L(\l)=L(\l)\ (\p\tp \textrm{id})\D(J^-). \label{j-}
\ee
From the asymptotics as $\lambda \to -\infty$ we obtain a relation similar to (\ref{j-}) for $J^+$, so we conclude:
\ba
&&(\p\tp \textrm{id})\D'(X)\ L(\l)=L(\l)\ (\p\tp \textrm{id})\D(X)\ \Rightarrow  \cr
&&(\p\tp\p)\D'(X)\ R(\l)=R(\l)\ (\p\tp\p) \D(X)  \qquad X \in U_q(\mathfrak{sl}_2). \label{check}
\ea
The second of the equations above suggests that the $R$-matrix satisfies linear intertwining relations with the
co-products of the underlying quantum algebra.
In fact, such types of linear exchange relations (\ref{check}) may be used in order to
extract $R$-matrices associated to
particular quantum algebras (Yangian or affine $q$ deformed algebras) see e.g. \cite{jimbo, kulish}. Note that extra
linear exchange relations involving the affine part of the associated quantum affine algebra are needed in
order to
fully identify the relevant $R$-matrix.
Similar relations are obtained for the Yangian from the asymptotics of the FRT equation, but are left
for the interested reader as exercise. Although this is a
systematic and elegant means to solve the YBE, we shall not further pursue this issue in this article.

\no At this point, recall that
$\check{R}=\mathcal{P}R$, and consequently
\be
(\p\tp\pi)\D(X)\ \check{R}(\l)=\check{R}(\l)\ (\p\tp\pi) \D(X) \qquad X \in U_q(\mathfrak{sl}_2).
\ee
\no
This equation is very important since it shows that $\check{R}$ commutes
with the generators of $U_q(\mathfrak{sl}_2)$, in the co-product realization;
 $\check R$ is called for obvious reasons the quantum invariant $R$ matrix \cite{jimbo}.

\paragraph{Exercise 1:} Consider the $\check R$-matrix of the XXZ spin chain, and check explicitly that
the above commutation relations (\ref{check}) hold $\forall X \in U_q(\mathfrak{sl}_2)$.

\section{Algebraic Bethe ansatz}

\subsection{$U_q(\mathfrak{sl}_2)$ representations}

Let us briefly review here various representation of the $U_q(\mathfrak{sl}_2)$ algebra.
First, recall the $L$-matrix (principal gradation)
\be
L(\l)=
\begin{pmatrix} e^\l A-e^{-\l}D & (q-q^{-1})B \cr (q-q^{-1})C & e^\l D-e^{-\l}A \end{pmatrix}
\label{lgeneric}
\ee

The spin $s$ representation is $n=2s+1$ dimensional defined in terms of $n\times n$ matrices as:
\ba
A  & = & \sum_{k=1}^{n}q^{a_k}e_{kk} =D^{-1} \qquad\qquad (\sim q^{S^z})\cr
B & = & \sum_{k=1}^{n-1} \widetilde{C}_k e_{k+1k} \qquad\qquad\qquad (\sim  S^-)\cr
C & = & \sum_{k=1}^{n-1} \widetilde{C}_k e_{kk+1} \qquad\qquad\qquad (\sim  S^+),
\ea
\no where
\ba
a_k & = &  q^{{n+1\ov2}-k}, \qquad \widetilde{C}_k=\sqrt{[k_q][n-k]_q},\cr
	&& [k]_q={q^k-q^{-k}\ov q-q^{-1}}.
\ea

\paragraph{Exercise 1:} Prove that $A,\ B,\ C,\ D$ satisfy $U_q(SL_2)$.

\paragraph{The Heisenberg-Weyl group:}

let us first introduce the Heisenberg-Weyl group, defined by elements ${\mathbb X},\ {\mathbb Y}$ that satisfy
\be
{\mathbb X}\ {\mathbb Y}=q\ {\mathbb Y}\ {\mathbb X}.
\ee
All the entries of the $L$ matrix may be then expressed in terms of the Heisenberg-Weyl elements as follows:
\ba
 {\mathrm A}= {\mathrm D}^{-1} = {\mathbb X},
~~~{\mathrm B} ={1 \over q-q^{-1}}(q^{-s}{\mathbb X}^{-1} - q^{s}{\mathbb X}){\mathbb Y}^{-1},
~~~{\mathrm C} ={1 \over q-q^{-1}}(q^{-s}{\mathbb X} - q^{s}{\mathbb X}^{-1}){\mathbb Y}.
\label{terms}
\ea

Let us also focus for a moment on the
special case where $q$ is root of unity, i.e. $q^{p}=1,\
q=e^{i\mu},\ \mu={2k\pi \over p}$, where $k,\ p$ integers. In this
case the algebra admits a $p$ dimensional representation, known as
the cyclic representation \cite{roar}. More specifically, one more restriction is applied
so one may
obtain a representation with no highest (lowest) weight
\be
{\mathbb X}^p={\mathbb Y}^p=1 \nonumber
\ee
then the generators ${\mathbb X},\ {\mathbb Y}$ may be
expressed as $p$ dimensional matrices
\be {\mathbb X} = \sum_{k=1}^p q^{-k}\
e_{kk},~~~~ {\mathbb Y}=\sum_{k=1}^{p-1}e_{k\ k+1} + e_{p1}. \label{cyclic}
\ee

\paragraph{Sine-Gordon and Liouville models:} in what
follows we shall briefly review how the lattice sine-Gordon \cite{izko} and
Liouville models \cite{fati} are obtained in a natural way from the XXZ $L$
matrix. Also the $q$ harmonic oscillator realization will be obtained
from the generalized XXZ form. The generators ${\mathbb X}$ and ${\mathbb Y}$ may be
associated with an infinite dimensional representation in terms of
some lattice `fields'.
\no Consider ${\mathbb Z}={\mathbb X}{\mathbb Y}$, then ${\mathbb Z}$ also satisfies
\be
{\mathbb X}\ {\mathbb Z}=q\ {\mathbb Z}\ {\mathbb X}.
\ee
\no Parametrizing
\be
{\mathbb X}_n=e^{-i\Phi_n}, \qquad {\mathbb Z}_n=e^{i\Pi_n},
\ee
\no where apparently $\Phi_n,\ \Pi_n$ are canonical
\be
[\Phi_n,\ \Pi_m]=i\m\d_{nm}, \qquad q^{s-1/2}=-i m.
\ee
The parameter $s$ of the representation is associated to the mass scale of the system. Also
by multiplying by $i m \sigma^x$ (we are allowed to multiply with $\sigma^x$ because this leaves the XXZ $R$ matrix
invariant) one obtains the lattice sine-Gordon L matrix \cite{izko}
\be
L_{an}^{S.G}=
\begin{pmatrix}
h_+(\phi_n)e^{i\Pi_n} & -2im\sinh(\l+i\Phi_n) \cr -2im\sinh(\l-i\Phi_n) & h_-(\Phi_n)e^{-i\Pi_n}
\end{pmatrix},
\ee
\no where
\be
h_{\pm}(\Phi_n)=1+m^2e^{\pm 2i\Phi_n+i\m}.
\ee
Consider
also the following limiting process \cite{fati} \be  i\Phi_{n} \to
i\Phi_{n} +c, ~~~~ \mu \lambda \to \mu \lambda +c, ~~~~m \to 0,
~~~~e^{-c} \to \infty, ~~~~m^2 e^{-2c} \to \alpha^{2} \label{lim}
\ee one obtains the lattice Liouville $L$ matrix
\be
L_{an}^{Lv}(\lambda) = \left(
\begin{array}{cc}
e^{i \Pi_{n}}   &\alpha e^{-\mu \lambda -i\Phi_{n}}\\
2 \alpha \sinh(\mu \lambda -i\Phi_{n})&h(\Phi_{n})e^{-i\Pi_{n}}\\
\end{array}\right )  \label{liou}
\ee
\be h(\Phi_{n})= 1 +\alpha^2 e^{-2i\Phi_{n} +i\mu}. \nonumber\ee
The interesting observation is that the entailed $L$ operator
(\ref{liou}) has a non trivial spectral ($\lambda$) dependence a
fact that allows the application of Bethe ansatz techniques for
the derivation of the spectrum (see also \cite{fati}).

The classical limit of the aforementioned $L$ matrices gives the corresponding
classical Lax operators
satisfying the zero curvature condition, and giving rise to the
classical equations of motion of the relevant models. Let us briefly review the
connection between the quantum (lattice) versions and the
classical sine-Gordon and Liouville models. Consider the following
classical limit \cite{izko, fati}, the spacing $\alpha \to 0$, set
$\mu \to h \mu$ such that ${1 \over h}[ , ] \to \{ , \}$, and
\be
\Phi_n \to {\beta \over 2} \phi(x) -{\pi \over 2}, ~~~~~\Pi_n \to
\alpha {\beta \over 4} \pi(x), ~~~~ -4im \to \alpha  \tilde m,
~~~\mu\lambda \to u +{i\pi \over 2}  \label{clas}
\ee
$\tilde m$ is the continuum mass and $\beta$ corresponds to the coupling
constant of the sine Gordon model, and for the Liouville model we
set ${\beta \over 2} =1$, following the normalization of
\cite{fati}. Bearing in mind the expressions above we
obtain as $\alpha \to 0$
\be
L(u) = 1 - \alpha {\mathrm U}(u) +{\cal O}(\alpha^2)
\ee
then the
quantities ${\mathrm U}(u)$ written below provides the Lax operator for
the classical continuum counterparts of the lattice sine-Gordon and
Liouville models. More precisely for the sine Gordon model:
\ba
&& {\mathrm U}(u) = {1\over 2} \left(
\begin{array}{cc}
-i {\beta \over 2}\pi(x)   &-\tilde m  \sinh(u+i{\beta \over 2}\phi(x))\\
\tilde m \sinh(u-i{\beta \over 2}\phi(x))  &   i{\beta \over 2}\pi(x)  \\
\end{array} \right), \label{sgc}
\ea
whereas for the Liouville
model the Lax operator reads
\ba
{\mathrm U}(u)= {1\over 2} \left(
\begin{array}{cc}
-i\pi(x)   & -2 e^{-u -i\phi(x)} \\
4 \sinh(u -i\phi(x)) &i\pi(x)\\
\end{array}\right ).  \label{liouc}
\ea
The Lax operators satisfy the classical analogue of the fundamental relation \cite{ftbook}. More precisely,
${\mathrm U}$ satisfies classical linear exchange relations described in \cite{ftbook}.
We shall not further discuss this topic here
given that is beyond the intended scope of the present article.

Similar limiting process to (\ref{lim}) leads to the $q$-harmonic
oscillator $L$ matrix starting from (\ref{lgeneric}) (see also \cite{doikoucyclic}). In
fact, by simply multiplying the Liouville $L$ matrix with an
anti-diagonal matrix we obtain
the following
\be
L_{an}(\lambda) =\left(\begin{array}{cc}
e^{\mu \lambda}V_{n} - e^{-\mu \lambda}V_{n}^{-1} &a_{n}^{+} \\
       a_{n}                      &-e^{-\mu \lambda}V_{n} \\ \end{array} \right ) \label{l2}
\ee
where the operators
$a_{n},\ a_{n}^{+},\ V_{n}$ are expressed in terms of ${\mathbb X}_{n},\
{\mathbb Y}_{n}$ as
\be
V_{n}= {\mathbb X}_{n}, ~~~~a_n^+ = ({\mathbb X}_{n}^{-1}
-q{\mathbb X}_{n}){\mathbb Y}_{n}^{-1}, ~~~a_{n}= {\mathbb Y}_{n}\ {\mathbb X}_{n} \ee and they satisfy
the $q$ harmonic oscillator algebra i.e. \be a_{n}^{+}\ a_{n} =
1-q V_{n}^{2}, ~~~~a_{n}\ a_{n}^{+} = 1-q^{-1}V_{n}^2.
\label{qalg}
\ee

\subsection{Algebraic Bethe ansatz}

Having introduced all the necessary algebraic setting we are now in a position to describe the algebraic
Bethe ansatz
method.
This can be basically applied for representations of Lie and deformed Lie algebras with highest (lowest)
weight.
For representations with no highest (lowest) weight the method can be applied with certain modifications,
which however will not
be discussed here (see e.g.\cite{FT2}). We shall extract below the spectrum and Bethe ansatz equations for the whole
hierarchy of the spin $s$
representations
of $U_q(\mathfrak{sl}_2)$.

The main objective within QISM is the diagonalization of the transfer matrix.
This will be achieved
by means of the
algebraic Bethe ansatz method \cite{FT, faddeev1, faddeev2}. We shall essentially exploit
the exchange relations emanating
from the fundamental algebraic relation
(\ref{rtt}) in order to determine the spectrum of the transfer matrix as well as
the corresponding eigenstates.
Recall that the transfer matrix is given by
\be
t(\l)=\textrm{Tr}_0[L_{0N}(\l)\cdots L_{01}(\l)]=\textrm{Tr}_0T_0(\l),
\ee
\no where
\be
L_{0n}(\l)=
\begin{pmatrix}
\sinh(\l+{i\m\ov2}+i\m J^z) & \sinh i\m J^- \cr \sinh i\m J^+ & \sinh(\l+{i\m\ov2}-i\m J^z)
\end{pmatrix}.
\ee
\no
The first step
is to determine a reference state  also called the ``pseudo-vacuum'' in the anti-ferromagnetic case.
Let $|\omega\rangle$ be the state annihilated
by $J^+$ and $|\Omega\rangle$ be the tensor product of $N$ such states:
\ba
&& J_n^+|\omega\rangle_n = 0\cr
&& |\Omega\rangle = |\omega\rangle\tp\cdots\tp|\omega\rangle.
\ea
\no Applying then $T(\l)$ to $|\Omega\rangle$ we get rid of the $C$'s, since
they annihilate the state
\ba
T(\l)|\Omega\rangle & = &
\begin{pmatrix} A_N & B_N \cr C_N & D_N \end{pmatrix} \cdots
\begin{pmatrix} A_1 & B_1 \cr C_1 & D_1 \end{pmatrix} |\omega\rangle\tp\cdots\tp|\omega\rangle\cr
& = & \begin{pmatrix} A_N & B_N \cr 0 & D_N \end{pmatrix} \cdots
\begin{pmatrix} A_1 & B_1 \cr 0 & D_1 \end{pmatrix} |\omega\rangle\tp\cdots\tp|\omega\rangle\cr
& = & \begin{pmatrix} A_N\cdots A_1 & \mathcal{B} \cr 0 & D_N\cdots D_1 \end{pmatrix} |\Omega\rangle\cr
& = & \begin{pmatrix} \mathcal{A} & \mathcal{B} \cr 0 & \mathcal{D} \end{pmatrix} |\Omega\rangle.
\ea
\no The exact form of $\mathcal{B}$ is not required, since we are going to
trace over the monodromy matrix, so only $\mathcal{A}+\mathcal{D}$ is needed. However, the
action of $A_n$ (or $D_n$ in the same spirit) on $|\omega\rangle$ is known, and is just
\be
\sinh(\l+{i\m\ov2}\pm i\m J_n^z)|\omega\rangle_n=\sinh(\l+{i\m\ov2} \pm i\m s)|\omega\rangle_n.
\ee
\no Thus we conclude that the action of $\mathcal{A}$
and $\mathcal{D}$ on $|\Omega\rangle$ is
\ba
\mathcal{A}|\Omega\rangle & = & \sinh\left(\l+{i\m\ov2}+i\m s\right)^N|\Omega\rangle\cr
\mathcal{D}|\Omega\rangle & = & \sinh\left(\l+{i\m\ov2}-i\m s\right)^N|\Omega\rangle.
\ea
\no The action of the transfer matrix on our pseudo-vacuum is then known
\ba
t(\l)|\Om\rangle & = & (\mathcal{A}+\mathcal{D})|\Omega\rangle\cr
& = & \left(\sinh\left(\l+{i\m \ov 2}+i\m s\right)^N+\sinh\left(\l+{i\m \ov 2}-i\m s\right)^N\right)
|\Omega\rangle.
\ea

The next step is to make the following ansatz for a general Bethe state $|\Psi\rangle$:
\be
|\Psi\rangle = \mathcal{B}(\l_1)\cdots \mathcal{B}(\l_M)|\Om\rangle.
\ee
\no We would like to find the action
of $\mathcal{A}+\mathcal{D}$ on $|\Psi\rangle$. Since we already know how $\mathcal{A}$
and $\mathcal{D}$ act on $|\Om\rangle$, we only need to determine the exchange
relations between $\mathcal{A},\ \mathcal{D}$ and $\mathcal{B}$.

As discussed the monodromy matrix satisfies (\ref{rtt}), with the $R$-matrix being the XXZ matrix:
\be
R_{12}(\l)=
\begin{pmatrix} \sinh(\lambda + i\mu) &0 &0 &0 \cr 0& \sinh\lambda & \sinh i \mu &0 \cr
0& \sinh i\mu & \sinh \lambda &0 \cr 0&0 &0 & \sinh(\lambda +i \mu)
\end{pmatrix}.
\ee
\no After some algebra, the fundamental algebraic relation gives the commutation
relations between $\mathcal{A},\mathcal{D}$ and $\mathcal{B}$. For example, one derives
($\delta = \lambda_1 - \lambda_2$):
\ba
[\mathcal{A}(\l_1),\ \mathcal{A}(\l_2)] & = &  [\mathcal{B}(\l_1),\ \mathcal{B}(\l_2)] =
 [\mathcal{C}(\l_1),\ \mathcal{C}(\l_2)]\cr
a(\d)\mathcal{B}(\l_1)\mathcal{A}(\l_2) & = & b(\d)\mathcal{A}(\l_2)\mathcal{B}(\l_1)
 + c\mathcal{B}(\l_2)\mathcal{A}(\l_1)\cr
a(\d)\mathcal{B}(\l_2)\mathcal{D}(\l_1) & = & b(\d)\mathcal{D}(\l_1)\mathcal{B}(\l_2)
+ c\mathcal{B}(\l_1)\mathcal{D}(\l_2).
\ea
\no The last terms on the RHS of the last two equation are ``unwanted''.
Acting with $(\mathcal{A}(\l)+\mathcal{D}(\l))$ on $|\Psi\rangle$ we have
\ba
(\mathcal{A}(\l)+\mathcal{D}(\l))|\Psi\rangle & = & (\mathcal{A}(\l)+\mathcal{D}(\l))
\mathcal{B}(\l_1)\cdots \mathcal{B}(\l_M)|\Om\rangle\cr
 & = & \left[{a(-\d_1)\ov b(-\d_1)}\cdots{a(-\d_M)\ov b(-\d_M)}\mathcal{B}(\l_1)\cdots
 \mathcal{B}(\l_M)\mathcal{A}(\l) + (\cdots)\right.\cr
 &  & \left.+ {a(\d_1)\ov b(\d_1)}\cdots{a(\d_M)\ov b(\d_M)}\mathcal{B}(\l_1)\cdots
 \mathcal{B}(\l_M)\mathcal{D}(\l) + (\cdots)\right]|\Om\rangle,
\ea
\no where $\delta_i = \l-\l_i$, and the $(\cdots)$ stand for the ``unwanted'' terms.
One sees that if these terms vanish, then $|\Psi\rangle$ is an
eigenstate of the transfer matrix, with known eigenvalues. Since we know how $\mathcal{A}$
and $\mathcal{D}$ act on $|\Om\rangle$, we may write then the above equation as
\ba
 && t(\lambda) |\Psi \rangle = \left[\prod_{i=1}^M{\sinh(\l-\l_i-i\m)\ov\sinh(\l-\l_i)}\sinh(\l+i\m s+{i\m\ov2})^N\right.\cr
 && + \prod_{i=1}^M{\sinh(\l-\l_i+i\m)\ov\sinh(\l-\l_i)}\sinh(\l-i\m s+{i\m\ov2})^N\cr
  & & +\left.(\cdots\cdots)\right]|\Psi\rangle.
\ea
\no It is therefore relevant to examine the conditions for the unwanted terms to vanish.
In fact, this is
merely true for some values of $\l$, which are denoted by $\l_i$ and are called ``Bethe
roots'', and satisfy the following set of equations
\be
\left({\sinh(\l_i+i\m s)\ov \sinh(\l_i-i\m s)}\right)^N = \prod_{i\neq j}^M{\sinh(\l_i-\l_j+i\m)\ov
\sinh(\l_i-\l_j-i\m)},
\ee
\no which are called \emph{Bethe Ansatz Equations} (BAE). As long as $\lambda_i$'s satisfy the BAE
the unwanted terms vanish. Moreover,
these equations guarantee the analyticity of the eigenvalues, and provide all the physical information
regarding the considered system.

\paragraph{Exercise 2:} Work out the details of the procedure for the XXZ model and
convince yourself that the unwanted terms really do vanish.

Having obtained the eigenvalues of the transfer matrix, which we now denote as $\L(\l)$,
one can obtain the energy and momentum eigenvalues of the system. The energy is
proportional to (see also \cite{faddeev1, faddeev2}) (we focus now on the spin 1/2 representation)
\be
\left.E\propto {d\ov d\l}\ln \L(\l)\right|_{\l=0},
\ee
\no and for the XXZ spin-1/2 model is found to be
\be
E=-{1\ov2\p}\sum_{j=1}^M{\m\sinh i\m\ov \sinh(\l_j+{i \m \over 2})\sinh(\l_j-{i\m \over 2})},
\ee
\no while the momentum is proportional to $\ln \L(0)$ and for the XXZ spin-1/2 model is
found to be
\be
P=-\sum_{j=1}^M \ln{\sinh(\l_j+{i\m\over 2})\ov\sinh(\l_j-{i\m\over 2})}.
\ee
\no Finally, the Bethe states are $\mathfrak{sl}_2$ highest weight states and are also
eigenstates of $J^z$ with eigenvalue
\be
J^z={N\ov2}-M.
\ee

\section{Reflection equation and open boundaries}

\subsection{The reflection equation}

So far we have just considered systems with periodic boundary conditions. To incorporate generic
boundary conditions that still preserve integrability we have to deal with another
quadratic algebra called the reflection algebra.
In this context
open spin chain like systems my be also considered, that is spin chains
with non trivial boundaries attached at their edges (see e.g.
\cite{cherednik, sklyanin, kusk}).
The starting point is to introduce a description of scattering of particles on a boundary, compatible with
the bulk consistency relations discussed in Lecture 3.
The boundary scattering is described by the so called reflection matrix, which satisfies the basic equation
called \emph{reflection equation} (RE), or \emph{boundary Yang-Baxter} \cite{cherednik, sklyanin}
equation
\be
R_{12}(\l_1-\l_2)K_1(\l_1)R_{21}(\l_1+\l_2)K_2(\l_2)=K_2(\l_2)R_{12}(\l_1+\l_2)K_1(\l_1)R_{21}(\l_1-\l_2),
\ee
\no where the $R$-matrix obeys the Yang-Baxter equation introduced in earlier lecture. The $K$-matrix
contains all the information about the reflection of the particle on the boundary.
Graphically,  the $R$ and $K$ matrices are represented as

\vspace{1cm}

\begin{center}
\begin{picture}(1,1)
	\put(-100,30){\line(1,-1){60}} \put(-100,-30){\line(1,1){60}}
	\put(-108,-35){$1$} \put(-52,-35){$2$}
	\put(0,0){\textrm{and}}
	\put(70,30){\line(0,-1){60}}
	\multiput(70,30)(0,-8){8}{\line(-1,-1){10}}
	\put(100,30){\line(-1,-1){30}} \put(70,0){\line(1,-1){30}}
\end{picture}
\end{center}

\vspace{1.5cm}

\no respectively. Using this graphical representation, the reflection equation
can be represented as

\vspace{1cm}

\begin{center}
\begin{picture}(1,1)
	\put(-100,30){\line(0,-1){60}}
	\multiput(-100,30)(0,-8){8}{\line(-1,-1){10}}
	\put(-60,25){\line(-2,-1){40}} \put(-100,5){\line(1,-1){30}}
	\put(-84,28){\line(-1,-3){16}} \put(-100,-20){\line(1,-1){30}}
	\put(-84,33){$2$} \put(-60,30){$1$}
	\put(-20,2){\line(1,0){10}}
	\put(-20,-2){\line(1,0){10}}
	\put(60,30){\line(0,-1){60}}
	\multiput(60,30)(0,-8){8}{\line(-1,-1){10}}
	\put(90,30){\line(-1,-1){30}} \put(60,0){\line(1,-2){20}}
	\put(90,15){\line(-1,-1){30}} \put(60,-15){\line(1,0){40}}
	\put(85,32){$2$} \put(95,15){$1$}
\end{picture}
\end{center}

\vspace{1.5cm}
In a number of situations the ``reflection'' matrix $K$ will indeed encapsulate boundary conditions
on the spin chain derived from the corresponding transfer matrices. The situation may be subtler in other instances,
i.e. dynamical reflection algebras, but we shall not further comment on these cases.

We can now find solutions of the reflection equation of the form
\be
K(\l)=
\begin{pmatrix} \a & \b \cr \g & \d \end{pmatrix},
\ee
\no where $\a,\ \b,\ \g,\ \d$ are c-numbers. This way, one may obtain several
c-numbers solutions of the reflection equation. For instance the generic non-diagonal solution for the
XXZ (sine-Gordon model)
found in \cite{GZ, DVGR} in the homogeneous gradation
\be
K(\lambda)=
\begin{pmatrix} \sinh(-\lambda + i \xi)e^{\lambda} & \kappa \sinh (2 \lambda) \cr \kappa \sinh (2 \lambda)
 & \sinh(\lambda + i \xi)e^{-\lambda} \end{pmatrix}.
\ee
The $K$ matrix in the principal gradation may by obtained via a gauge transformation
\be
K^{(p)}(\lambda) = V(-\lambda)\ K^{(h)}(\lambda)\ V(-\lambda)
\ee

However, one would
also like to find solutions where the elements of $\mathbb{K}$ are
now operators and not c-numbers. These type of operatorial solutions are of the generic from
\cite{sklyanin}
\be
\mathbb{K}(\l)=L(\l)\ K^-(\l)\ L^{-1}(-\l),
\ee
\no where $K^-(\l)$ is a c-number solution and $L(\l)$ satisfies (\ref{RLL})
We also define the corresponding modified monodromy matrix \cite{sklyanin}
\be
\mathbb{T}(\l)=T(\l)\ K^-(\l)\ T^{-1}(-\l),
\ee
\no where $T(\l)$ is the already known monodromy matrix
\be
T(\l)=L_{0N}(\l)\cdots L_{01}(\l)
\ee
\no Moreover, we introduce $K^+$ as
\be
K^+(\l)=MK^t(-\l-i\r),
\ee
\no where $\r$ is the crossing parameter and for the $\mathfrak{gl}_n,\ U_q(\mathfrak{gl}_n)$
cases is $\rho = {n \over 2}$, $^t$
stands for transposition,
and $K$ is a solution of the reflection equation. Also $M$ is in general a diagonal matrix such that
\be
\Big [ R_{12}(\lambda),\ M_1 M_2\Big ] = 0
\ee
and for the XXZ model in particular
\ba
M &=& {\mathbb I}~~~~\mbox{principal gradation}, \cr
M &=& \mbox{diag}(q,\ q^{-1}) ~~~~~\mbox{homogeneous gradation}. \label{mm}
\ea
This way,
we are able to define the open transfer matrix as
\be
t(\l)=\textrm{Tr}_0[K^+(\l)\ \mathbb{T}(\l)]=\textrm{Tr}_0[K^+(\lambda)\ T(\l)\ K^-(\l)\ T^{-1}(-\l)].
\ee
\no One can show then that the integrability
condition, that is
\be
[t(\l),\ t(\l')]=0.
\ee

\subsection{Solutions of the RE and $B$-type braid group}

Just as in the case of the closed spin chains, we can systematically search for solutions of
the reflection equation by exploiting the similarity of the latter with the $B$-type braid
group (see e.g \cite{cher, male, mawo, doma} and references within).
\\
\\
{\bf Definition 8.1.} {\it The $B$-type braid group consists of $N-1$ generators $g_i$, which
satisfy the already known braid relations}
\ba
g_i\ g_{i+1}\ g_i & = & g_{i+1}\ g_i\ g_{i+1}, \qquad i \in\{ 1, \ldots, N-2\},\cr
[g_i,\ g_j] & = & 0, \qquad |i-j|>1,
\ea
\no {\it plus an additional generator $g_0$ which satisfies}
\ba
g_0\ g_1\ g_0\ g_1 & = & g_1\ g_0\ g_1\ g_0, \cr
[g_i,\ g_0] & = & 0, \qquad i>1.
\ea
\no The similarity of the latter relation with the `modified' reflection equation
\be
\check{R}_{12}(\lambda_1-\lambda_2)K_1(\lambda_1)\check{R}_{12}(\lambda_1+\lambda_2)K_1(\lambda_2)=
K_1(\lambda_2)\check{R}_{12}(\lambda_1+\lambda_2)K_1(\lambda_1)\check{R}_{12}(\lambda_1 -\lambda_2)
\ee
\no suggests that finding representations of the $B$-type braid group is equivalent to finding
solutions of the RE. However the $B$-type braid group is too big to be physical, therefore we shall restrict our attention to
certain quotients, which will be introduced below.
\\
\\
{\bf Definition 8.2.} {\it The affine Hecke algebra is defined by generators
 $g_i,\ g_0$ that satisfy the $B$-type conditions plus}
\be
(g_i-q)(g_i+q^{-1})=0,\qquad \textrm{(Hecke condition)}.
\ee
\\
{\bf Definition 8.3.} {\it The Cyclotomic algebra is a quotient of the affine Hecke algebra obtained
by imposing the
extra constraint}
\be
\prod_{i=1}^n(g_0-\g_i)=0,
\ee
\no where $\g_i$ are free parameters.
\\
\\
{\bf Definition 8.4.} {\it The $B$-type Hecke algebra $B_N(q,\ Q)$ is a quotient of the affine Hecke algebra
satisfying the extra condition:}
\be
(g_0-Q_1)(g_0-Q_2)=0,
\ee
\no {\it where usually $Q_2$ is taken to be $-Q_1^{-1}$}.

For the next definition it is convenient to introduce the following alternative generators:
\ba
U_0 & = & g_0-Q,\cr
U_i & = & g_i-q,
\ea
\\
{\bf Definition 8.5.}
{\it The boundary Temperley-Lieb (blob) algebra ${\cal B}_N(q,\ Q)$ with generators $U_i,\ U_0$ is a quotient
of the $B$-type
Hecke algebra with exchange relations}
\no
\ba
&& U_{i\pm 1}\ U_i\ U_{i\pm1}=U_{i\pm1}\cr
&& U_i^2=-(q+q^{-1})U_i\cr
&& U_1\ U_0\ U_1 = \k U_1, \qquad \k=qQ^{-1}+q^{-1}Q\cr
&& U_0^2=-(Q+Q^{-1})U_0.
\ea

\paragraph{Graphical Representation of the boundary Temperley-Lieb algebra:} recall the
generator $U_i$ is defined as

\vspace{2cm}

\begin{center}
\begin{picture}(1,1)
	\put(0,0){\line(1,0){120}} \put(0,0){\line(-1,0){120}}
	\put(0,60){\line(1,0){120}} \put(0,60){\line(-1,0){120}}
	\put(-90,60){\line(0,-1){60}} \put(90,60){\line(0,-1){60}}
	\put(-65,25){$\cdots$} \put(55,25){$\cdots$}
	\put(-90,65){$1$} \put(80,65){$N$}
	\put(-30,65){$i$} \put(15,65){$i+1$}
	\put(0,60){\oval(60,40)[b]}
	\put(0,0){\oval(60,40)[t]}

\end{picture}
\end{center}

\no Depict also $U_0$ as

\vspace{2cm}

\begin{center}
\begin{picture}(1,1)
	\put(0,0){\line(1,0){120}} \put(0,0){\line(-1,0){120}}
	\put(0,60){\line(1,0){120}} \put(0,60){\line(-1,0){120}}
	\put(-90,60){\line(0,-1){60}} \put(90,60){\line(0,-1){60}}
	\put(-30,60){\line(0,-1){60}}
	\put(-90,65){$1$} \put(80,65){$N$}
	\put(-30,65){$2$} \put(15,65){$\cdots$} \put(15,25){$\cdots$}
	\put(-90,30){\circle*{5}}

\end{picture}
\end{center}

\no The relation $U_1\ U_0\ U_1=\k\ U_1$ is represented then as

\vspace{6.5cm}

\begin{center}
\begin{picture}(1,1)


	\put(-90,0){\line(1,0){80}} \put(-90,0){\line(-1,0){100}}
	\put(-90,60){\line(1,0){80}} \put(-90,60){\line(-1,0){100}}
	\put(-80,60){\line(0,-1){60}} \put(-20,60){\line(0,-1){60}}
	 \put(-55,25){$\cdots$}
	\put(-150,60){\oval(60,40)[b]}
	\put(-150,0){\oval(60,40)[t]}

	\put(-20,120){\line(0,-1){60}} \put(-180,120){\line(0,-1){60}}
	\put(-120,120){\line(0,-1){60}} \put(-80,120){\line(0,-1){60}}
	\put(-55,85){$\cdots$}
	\put(-180,90){\circle*{5}}

	\put(-90,120){\line(1,0){80}} \put(-90,120){\line(-1,0){100}}
	\put(-90,180){\line(1,0){80}} \put(-90,180){\line(-1,0){100}}
	\put(-80,180){\line(0,-1){60}} \put(-20,180){\line(0,-1){60}}
	 \put(-55,145){$\cdots$}
	\put(-180,185){$1$} \put(-30,185){$N$}
	\put(-120,185){$2$} \put(-80,185){$3$}
	\put(-150,180){\oval(60,40)[b]}
	\put(-150,120){\oval(60,40)[t]}

	\put(0,89){\line(1,0){10}}
	\put(0,86){\line(1,0){10}}


	\put(50,120){\oval(30,40)[t]}
	\put(50,60){\oval(30,40)[b]}
	\put(35,58){\line(0,1){65}}
	\put(65,58){\line(0,1){65}}
	\put(35,90){\circle*{5}}

	\put(180,60){\line(1,0){60}} \put(180,60){\line(-1,0){100}}
	\put(180,120){\line(1,0){60}} \put(180,120){\line(-1,0){100}}
	\put(230,120){\line(0,-1){60}} \put(180,120){\line(0,-1){60}}
	 \put(200,85){$\cdots$}
	\put(120,120){\oval(60,40)[b]}
	\put(120,60){\oval(60,40)[t]}

\end{picture}
\end{center}
\no The closed line with the dot in the RHS of the graph above stands for the constant $\kappa$.

\paragraph{Exercise 1:} Consider the matrices
\be
U =
\begin{pmatrix} 0 &0 &0 &0\cr 0& -q & 1 &0 \cr 0& 1 & -q^{-1} &0 \cr 0 &0 &0 &0 \end{pmatrix},
\qquad e =
\begin{pmatrix} -Q^{-1} & c \cr c^{-1} & -Q \end{pmatrix}. \label{umatrix}
\ee
\no Show that they provide a representation of the boundary Temperley-Lieb algebra,
$\pi: {\cal B}_N(q,\ Q) \hookrightarrow \mbox{End}(({\mathbb C}^2)^{\otimes N})$
\ba
&& \pi(U_i)= {\mathbb I} \otimes \ldots \otimes {\mathbb I} \otimes \underbrace{U}_{i,\ i+1}
\otimes {\mathbb I} \otimes \ldots \otimes {\mathbb I} \cr
&& \pi(U_0) = \underbrace{e}_{1^{st}} \otimes {\mathbb I} \otimes \ldots \otimes {\mathbb I}.
\ea

\paragraph{Exercise 2:} Take the definition of the transfer matrix for the open spin
chains with boundaries. Suppose also that
\ba
\hat{R}_{i i+1} & = & a(\l)U_i+b(\l)\cr
K^- & = & x(\l)\mathbb{I}+y(\l)U_0\cr
K^+ &=& M ~~~~\mbox{and} ~~~~tr_0[M_0 \check R_{N0}(0)] \propto {\mathbb I}
\ea
\no where
\ba
y(\l) & = & 2\sinh i\m\ \sinh2\l\cr
x(\l) & = & (Q+Q^{-1})\cosh(2\l+i\m)-\cosh(2i\m\g)-\k\cosh(2\l),
\ea
\no with $Q=ie^{i\m m}, \g$ free parameters. Show that (see also \cite{doma, doikouh})
\be
\left.H = {d\ov d\l}t(\l)\right|_{\l=0} \propto \sum_{i=1}^N U_i +c_1 U_0 + c_2.
\ee

\subsection{The $U_q(\mathfrak{sl}_2)$ invariant spin chain}

We shall now deal with the $U_q(\mathfrak{sl}_2)$ invariant open spin chain \cite{kusk1}
(see also relevant discussion on quantum symmetries in general, and associated properties
in \cite{dede}). We shall exhibit the
symmetry of the spin chain following a logic quite
different from the one presented in the original work of \cite{kusk1}.
Consider the simplest case of boundary conditions, that is $K^- \propto {\mathbb I}$ and $K^+ =M$ (\ref{mm}).

We shall focus here on the
homogeneous gradation.
Recall that in this case the $L$ matrix may be decomposed in upper/lower triangular matrices giving rise
to upper
lower Borel sub-algebras in a natural way.
The tensor representation of the reflection equations is
\be
{\mathbb T}(\lambda) = T(\lambda)\ \hat T(\lambda).
\ee
Recall the linear algebraic relations (\ref{check}). It is then clear that
\be
(\pi \otimes \mbox{id}^{\otimes N})\Delta^{'(N+1)}(X)\ T(\lambda) = T(\lambda)\ (\pi \otimes
\mbox{id}^{\otimes N})\Delta^{(N+1)}(X),
~~~X \in U_q(\mathfrak{sl}_2).
\ee
Also it is easy to verify, recalling the form of $\hat T$, that
\be
(\pi \otimes \mbox{id}^{\otimes N})\Delta^{(N+1)}(X)\ \hat T(\lambda) = \hat T(\lambda)\
(\pi \otimes \mbox{id}^{\otimes N})
\Delta^{'(N+1)}(X).
\ee
Combining the relations above and bearing in mind the form of ${\mathbb T}$ for the
particular choice of boundaries
($K \propto {\mathbb I}$)
we conclude:
\be
(\pi \otimes \mbox{id}^{\otimes N})\Delta^{'(N+1)}(X)\ {\mathbb T}(\lambda) = {\mathbb T}(\lambda)\
(\pi \otimes \mbox{id}^{\otimes N})\Delta^{'(N+1)}(X),
~~~X \in U_q(\mathfrak{sl}_2). \label{basic2}
\ee
Let us fist consider $X= J^z$ then the relation above becomes
\be
({\sigma^z \over 2} \otimes {\mathbb I} + {\mathbb I} \otimes \Delta^{(N)}(J^z))\ {\mathbb T}(\lambda) =
{\mathbb T}(\lambda)\ ({\sigma^z \over 2} \otimes {\mathbb I} + {\mathbb I} \otimes \Delta^{(N)}(J^z)). \label{jz}
\ee
Multiply both sides with $M \otimes {\mathbb I}$ and take the trace over the auxiliary space then obtain
\ba
&& [\Delta^{(N)}(J^z),\ tr_0\{M_0 {\mathbb T}_0(\lambda)\}] =
{1\over 2} (tr_0\{\sigma_0^z{\mathbb T}_0(\lambda)\} - tr_0\{{\mathbb T}_0(\lambda)\sigma_0^z\}) =0 \cr
&& \Rightarrow [\Delta^{(N)}(J^z),\ t(\lambda)] =0.
\ea
Equation (\ref{jz}) can be equivalently expressed as
\be
q^{\pm{\sigma^z \over 2}} \otimes \Delta^{(N)}(q^{\pm J^z})\ {\mathbb T}(\lambda) =
{\mathbb T}(\lambda)\ q^{\pm{\sigma^z \over 2}} \otimes \Delta^{(N)}(q^{\pm J^z}), \label{alter}
\ee
and this form is more convenient in what follows.

Consider now $X =J^{\pm}$ then (\ref{basic2}) becomes
\ba
\Big (q^{{\sigma^z \over 2}} \otimes \Delta^{(N)}(J^{\pm}) + \sigma^{\pm} \otimes
\Delta(q^{-J^z})\Big )\ {\mathbb T}(\lambda)=
{\mathbb T}(\lambda)\ \Big (q^{{\sigma^z \over 2}} \otimes \Delta^{(N)}(J^{\pm}) +
\sigma^{\pm} \otimes \Delta(q^{- J^z}) \Big ).
\ea
Multiply both sides with $M q^{-{\sigma^z \over 2}} \otimes {\mathbb I}$ then the latter expression becomes
\ba
&&\Big [\Delta^{(N)}(J^{\pm}), \ M{\mathbb T}(\lambda)\Big ]= \cr
 && M {\mathbb T}(\lambda) q^{-{\sigma^z\over 2}} \sigma^{\pm}\otimes \Delta^{(N)}
 (q^{-J^z}) -M q^{-{\sigma^z\over 2}}\sigma^{\pm}
 \otimes \Delta(q^{-J^z}) {\mathbb T}(\lambda).
 \label{last}
\ea
After taking the trace over the auxiliary trace,
bearing in mind (\ref{alter}), and appropriately moving the elements within the trace we conclude
\be
\Big [\Delta^{(N)}(J^{\pm}),\ t(\lambda) \Big ] =0,
\ee
and with this we conclude our proof on the $U_q(\mathfrak{sl}_2)$ symmetry of the transfer matrix.
The proof can be easily generalized for any higher rank quantum algebra.

In general, special choice of $K$-matrix may suitably break down the symmetry, and depending on the
structure of the reflection matrix just part of the algebra
or particular combinations of the algebra elements may commute with the transfer matrix
(see e.g. \cite{done, dema, doikou1, doikouh}).
Although this is a particularly
interesting issue we shall not further discuss it here.

To obtain the local Hamiltonian we focus on the case where both auxiliary and quantum spaces are represented
by the fundamental representation of the considered Lie algebra,
that is $L(\lambda) \hookrightarrow R(\lambda)$.
The Hamiltonian in this case is obtained from the derivative of the transfer matrix, and has the universal form
in terms of the Temperley-Lieb generators (see also Exercise 2), recall (\ref{tl0})
\be
H \propto \sum_{i =1}^{N-1} U_i, \label{local1}
\ee
$U_i$ is any TL algebra representation. Here we are focusing on $U_q(\mathfrak{sl}_2)$ (XXZ model),
but the later expression
of the local Hamiltonian (\ref{local1}) is universal,
i.e. independent of the choice of the representation of the Temperley-Lieb algebra.
For the XXZ chain ($U_q(\mathfrak{sl}_2)$) in particular the Hamiltonian may be expressed in terms
of Pauli matrices as:
\ba
H \propto {1\over 2} \sum_{i=1}^{N-1}\Big (\sigma_i^x\sigma_{i+1}^x +\sigma_i^y\sigma_{i+1}^y
 +\cosh(i \mu)\sigma_i^z\sigma_{i+1}^z \Big ) +
{\sinh(i\mu)\over 2} \Big (\sigma_N^z - \sigma_1^z \Big ) - (N-1){\cosh(i\mu)\over 2} \cr
\ea
and it is manifestly $U_q(\mathfrak{sl}_2)$ invariant \cite{pasquiersaleur}.

We shall finally identify in a simple manner the quadratic Casimir
operators of $U_q(\mathfrak{sl}_2)$.
Focus for simplicity
on the case where $N=1$,
then the asymptotic behavior of ${\mathbb T}$
\be
{\mathbb T}^{\pm} = L^{\pm}\ \hat L^{\pm}
\ee
More precisely (recall $\hat L(\lambda) = L^{-1}(-\lambda)$)
\be
{\mathbb T}^{+} = \left(
  \begin{array}{cc}
    q^{{1\over 2}+c}q^{J^z} &(q-q^{-1}) J^- \\
    0 &  q^{{1\over 2}+c}q^{-J^z} \\
  \end{array}
\right) \  \left(
  \begin{array}{cc}
    q^{{1\over 2}-c}q^{J^z} & 0\\
    (q-q^{-1}) J^+          & q^{{1\over 2}-c}q^{-J^z} \\
  \end{array}
\right) \
\ee
and
\be
{\mathbb T}^{-} = \left(
  \begin{array}{cc}
    q^{-{1\over 2}-c}q^{-J^z} & 0\\
     (q^{-1}-q) J^+&  q^{-{1\over 2}-c}q^{J^z} \\
  \end{array}
\right) \  \left(
  \begin{array}{cc}
    q^{-{1\over 2}+c}q^{-J^z} &(q^{-1}-q) J^-  \\
    0& q^{-{1\over 2}+c}q^{J^z} \\
  \end{array}
\right) \
\ee
then the asymptotics of the transfer matrix provides the quadratic Casimir operators of $U_q(\mathfrak{sl}_2)$
\ba
&& t^{\pm} = tr\{M{\mathbb T}^{\pm}\} \Rightarrow \cr
&& t^+ \propto qq^{2J^z} +q^{-1}q^{2J^z} + (q-q^{-1})^2J^-J^+ \cr
&& t^- \propto q^{-1}q^{2J^z} +q q^{-2J^z} + (q-q^{-1})^2J^+J^-.
\ea
It is clear that for generic $N$ all algebra elements $X \to \Delta^{(N)}(X),\ X \in U_q(\mathfrak{sl}_2)$. It is thus clear
that the study of the asymptotic behavior of a spin chain may provide in a very simple and elegant manner
all the Casimir operators associated
to any Lie or deformed Lie algebra (see also \cite{doikoucyclic, doikousfetsos} for more details).

\paragraph{Exercise 3:} Consider the XXZ representation of the Temperley-Lieb algebra,
$\rho: T_N(q) \hookrightarrow \mbox{End}(({\mathbb C}^2)^{\otimes N})$ such that
\be
\rho(U_i) = {\mathbb I} \otimes \ldots \otimes {\mathbb I} \otimes \underbrace{U}_{i,\ i+1}
\otimes {\mathbb I} \otimes \ldots \otimes {\mathbb I} ,
\ee
where the matrix $U$ is defined in (\ref{umatrix});
consider also the spin ${1 \over 2}$ representation of $U_q(\mathfrak{sl}_2)$. Show that:
\be
[\rho(U_i),\ \Delta^{(N)}(x)] =0, ~~~~~x \in \{\sigma^{\pm},\ q^{\sigma^z\over 2}\},
\ee
i.e.
the $U_q(\mathfrak{sl}_2)$ algebra is central to the Temperley-Lieb algebra.
\\
\\
{\bf Acknowledgments}
\\ We are indebted to J. Avan and K. Sfetsos for valuable comments and suggestions on the manuscript.
A.D. and G.F. wish to thank the Physics Department of the
University of Bologna for kind hospitality.


\end{document}